\documentclass[twocolumn, amsmath, amssymb, floatfix, reprint, twocolumn, notitlepage, nofootinbib]{revtex4-1}
\pdfoutput=1
\usepackage[utf8]{inputenc}
\usepackage[english]{babel}
\usepackage[T1]{fontenc}
\usepackage{amsmath}
\usepackage{amssymb}
\usepackage{hyperref}
\usepackage{braket}
\usepackage[doipre={DOI:~}]{uri}

\usepackage{tikz}
\usepackage{lipsum}

\newcommand*\bx{\mathbf{x}}
\newcommand*\by{\mathbf{y}}
\newcommand*\bk{\mathbf{k}}
\newcommand*\hOmega{\Omega}

\begin{document}

\title{Few-particle scattering from localized quantum systems in spatially structured bosonic baths}

\author{Rahul Trivedi}
\affiliation{Max Planck Institut f\"ur Quantenoptik, Garching bei M\"unchen, 85748, Germany.}
\email{rahul.trivedi@mpq.mpg.de}
\thanks{Source code for the numerical simulations in this paper is available \href{https://github.com/rahultrivedi1995/scattering_high_dim_baths}{here}}
\author{Kevin Fischer}
\affiliation{Stanford University, Stanford, CA, 94305, USA.}
\author{Shanhui Fan}
\affiliation{Stanford University, Stanford, CA, 94305, USA.}
\author{Jelena Vuckovic}
\affiliation{Stanford University, Stanford, CA, 94305, USA.}

\begin{abstract}
Understanding the dynamics of localized quantum systems embedded in engineered bosonic environments is a central problem in quantum optics and open quantum system theory. We present a formalism for studying few-particle scattering in a localized quantum system interacting with an bosonic bath described by an inhomogeneous wave-equation. In particular, we provide exact relationships between the quantum scattering matrix of this interacting system and frequency domain solutions of the inhomogeneous wave-equation thus providing access to the spatial distribution of the scattered few-particle wave-packet. The formalism developed in this paper paves the way to computationally  understanding the impact of structured media on the scattering properties of localized quantum systems embedded in them without simplifying assumptions on the physics of the structured media.
\end{abstract}

\maketitle

\section{Introduction}
Non-classical light sources \cite{lodahl2015interfacing,ding2016demand,michler,senellart2017high,zhang2018strongly, kok2007linear,o2009photonic,roy2017colloquium,reiserer2015cavity,duan2010colloquium,sangouard2011quantum,nemoto2014photonic} are an essential resource in all optical quantum information processing and communication systems. Traditionally, such light sources have been designed by interfacing localized quantum systems (quantum dots, color centers etc.) with a bosonic bath (an electromagnetic structure such as an optical fiber, waveguide or optical cavity). The localized quantum system can modify the statistics and the entanglement structure of an incoming state propagating in the bosonic bath \cite{englund2005controlling,daveau2017efficient, pelton2002efficient,he2013demand, fischer2017signatures, hanschke2018quantum,metcalfe2010resolved,miao2019electrically, lukin2020spectrally}. This has opened up the possibility of generating high-fidelity single photon sources \cite{he2013demand, wang2019towards}, photon pair sources \cite{li2004all, harder2013optimized} as well as high-dimensional entangled photon states \cite{pichler2016photonic, xu2018generate} by engineering both the bath as well as the Hamiltonian of the localized quantum systems.

From a theoretical standpoint, a quantitative theory of the dynamics and scattering properties of this coupled system is required for designing any of the above-mentioned applications. If only the dynamics of the localized system is of interest, the usual approach is to trace out the Hilbert space of the bath to obtain a master equation for the density operator of the localized system \cite{carmichael2009open, gardiner2004quantum}. While traditionally the master equation is formulated for a one-dimensional homogeneous bath \cite{gardiner2004quantum}, it is easily extended to cases where the bath is described by an inhomogeneous 3D wave-equation \cite{asenjo2017atom, dung2002resonant}. However, a significant drawback of the master equation formalism is that it is not amenable to computing the state of the bath, although recent works have attempted to resolve this problem for 1D homogeneous baths \cite{kiilerich2019input}.

An alternative set of computational and analytical tools which allow us to calculate the bath's quantum state is provided by scattering theory \cite{taylor2006scattering}. Here, the localized system is treated as a scatterer for the quantum states that otherwise propagate in  the bath. Significant progress has been made on scattering theory for systems where a modal decomposition of the bosonic bath is easily accessible. In particular, there has been great success in developing techniques for analyzing the scattering matrices of the localized quantum system when the electromagnetic structure is a photonic waveguide \cite{fan2010input, xu2013analytic, xu2015input, trivedi2018few, trivedi2021optimal,fischer2017scattering, shi2015multiphoton, caneva2015quantum}, a cavity \cite{haroche1993cavity, trivedi2019photon, shi2013two} or a network of these components. There has also been progress towards understanding the interaction of localized quantum systems with 3D homogeneous baths \cite{liu2016quantum} and periodically structured baths \cite{orenstein2020two}. Computing the scattering of few-photon states in systems where the bosonic bath is describable as arbitrary spatially varying permittivity distributions remains an open problem. Such a framework, if it exists, can enable both analysis and design of more complex quantum photonic devices similar to how computational electromagnetic methods capable of simulating arbitrary permittivity distributions revolutionized the design of classical photonic devices \cite{miller1988selective, molesky2018inverse}.

In this paper, we present a framework for exactly calculating the scattering matrix of a localized quantum system interacting with a bosonic bath whose dynamics are described by an inhomogeneous wave equation. There are two key ingredients that go towards making this framework possible --- 
\begin{enumerate}
\item A second quantized description of the inhomogeneous bosonic bath in terms of both Langevin noise operators and plane-wave annihilation operators (Section \ref{sec:inh_bath}). Here, we build upon the formalism presented in Refs.~\cite{dung1998three, gruner1996green} for quantizing lossy frequency-dependent and inhomogeneous electromagnetic environments.
\item Calculation of few-particle scattering matrices that builds on this second quantized description (Section~\ref{sec:tls_scat}). We express the spatial and spectral dependence of the few-particle scattering matrices in terms of two quantities --- \emph{first}, the frequency-domain solutions of the inhomogeneous wave-equation describing the bath dynamics and \emph{second}, the time-ordered expectations of the localized system operators (Section \ref{sec:tls_scat}). While our general approach is applicable to arbitrary localized systems, we provide exact expressions for the single and two-particle scattering matrices for the paradigmatic example of a two-level localized system.
\end{enumerate}

\section{Quantization of an inhomogeneous bosonic bath}\label{sec:inh_bath}
\begin{figure}[b!]
\centering
\includegraphics[scale=0.35]{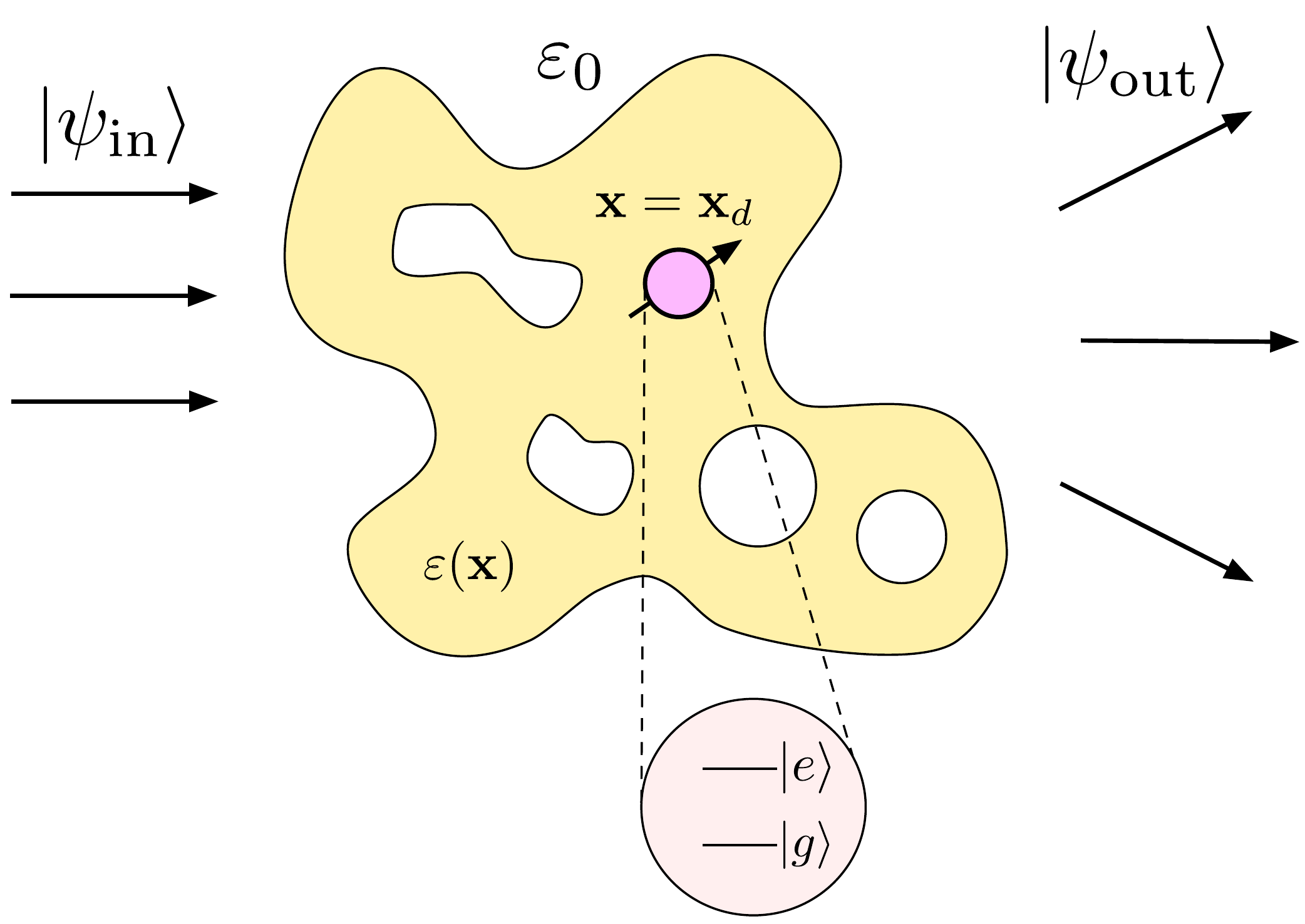}
\caption{\textbf{Schematic.} A localized two-level system interacting with an inhomogeneous bosonic bath described by a permittivity distribution $\varepsilon(\bx)$ which is asymptotically homogeneous ($\varepsilon(\bx) \to \varepsilon_0$ as $|\bx| \to \infty$). An incident few-particle state $\ket{\psi_\text{in}}$ propagates through the inhomogeneous bath and scatters off the localized system to produce an outgoing few-particle state $\ket{\psi_\text{out}}$.}
\label{fig:schematic}
\end{figure}

Figure \ref{fig:schematic} schematically depicts the systems under consideration comprising of a localized quantum system interacting with an inhomogeneous bosonic bath. We assume that the bosonic bath is classically described by the inhomogenous wave-equation
\begin{equation}\label{eq:wave_eq}
\nabla^2 \psi(\bx, t) = \varepsilon(\bx)\frac{\partial^2}{\partial t^2} \psi(\bx, t),
\end{equation}
where $\psi(\bx, t)$ is the time-dependent field variable and $\varepsilon(\bx)$ is the permittivity distribution describing the inhomogeneity in the bath. While we restrict ourselves to a simple scalar wave-equation model in this paper, an extension to a full vector-wave equation model is also possible. Furthermore, we will assume that the bosonic bath in isolation is lossless and consequently the permittivity distribution $\varepsilon(\bx)$ is real and positive. {\color{black}Furthermore, we assume that the inhomogeneous media is embedded in a background homogeneous media with permittivity $\varepsilon_0$ i.e.~$\varepsilon(\bx) \to \varepsilon_0$ as $|\bx|\to \infty$.}
\subsection{First quantization}
A quantum theory for such a bosonic bath can easily be developed by applying Dirac's quantization \cite{weinberg1995quantum} procedure on its classical Lagrangian $\mathcal{L}_\text{bath}[\psi]$
\begin{equation}
\mathcal{L}_\text{bath}[\psi] =\frac{1}{2} \int_{\mathbb{R}^3} \bigg( \varepsilon(\bx) \bigg|\frac{\partial \psi(\bx, t)}{\partial t}\bigg|^2 - |\nabla \psi(\bx, t)|^2\bigg) d^3 \bx.
\end{equation}
From this lagrangian, the canonically conjugate field to $\psi(\bx, t)$ is easily identified to be
\begin{align}\label{eq:conj_eq}
 \pi(\bx, t) = \varepsilon(\bx) \frac{\partial \psi(\bx, t)}{\partial t}.
 \end{align}
  Dirac's quantization then promotes both of these fields to operators, $\psi(\bx)$ and $\pi(\bx)$, with commutator $[\psi(\bx), \pi(\bx')] = i\delta^3(\bx - \bx')$. Furthermore, the dynamics of the quantum state of the bath is governed by the Hamiltonian
\begin{align}
H_\text{bath} = \int_{ \mathbb{R}^3} \bigg(\frac{\pi^2(\bx)}{2\varepsilon(\bx)} + \frac{1}{2}|\nabla \psi(\bx)|^2\bigg)d^3\bx
\end{align}
It can be readily verified that the Heisenberg equations of motion for the operators $\psi(\bx)$ and $\pi(\bx)$ reproduce the equations of motion for the classical fields as given by Eqs.~\ref{eq:wave_eq} and \ref{eq:conj_eq}.

\subsection{Second quantization}
{\color{black}While the first quantized description provides a complete quantum theory for the bosonic bath, it does not allow for an easy and physically meaningful description of incident and scattered quantum states. It is, therefore, of interest to perform second quantization and introduce particle annihilation and creation operators. We accomplish a second quantization of an inhomogeneous bosonic bath in two steps. \emph{First}, we consider a more general case of lossy inhomogeneous media and introduce a second quantization based on Langevin noise formalism. \emph{Second}, we treating lossless inhomogeneous media as a limiting case of lossy inhomogeneous media, we construct a second quantization in which an annihilation operator is associated with each possible plane-wave incident on the inhomogeneous media.}

\subsubsection{Langevin noise formalism}
In this subsection, we build upon the formalism introduced in Refs.~\cite{dung1998three, gruner1996green} which quantize the wave-equation in lossy inhomogeneous Kramer-Kronig dielectrics using a Langevin noise-operator approach. {\color{black}However, since we are fundamentally interested in lossless inhomogeneous baths, to make use of this formalism we fictitiously introduce loss in the bosonic bath, and consider a frequency-dependent permittivity distribution $\varepsilon_\alpha(\bx, \omega)$ given by
\begin{align}
\varepsilon_\alpha(\bx, \omega) = \varepsilon(\bx) + i\alpha(\omega),
\end{align}
where $\alpha(\omega) > 0$ is the loss in the permittivity distribution. For simplicity, we assume the loss to be position independent --- the permittivity distribution $\varepsilon_\alpha(\bx, \omega)$ is thus that of a lossy inhomogeneous permittivity distribution in a homogeneous background with permittivity $\varepsilon_0 + i\alpha(\omega)$. While we could work with a position dependent loss as well, since the objective is to study the limiting case of a lossless permittivity distribution, the position dependence of the loss does not matter.}

With the introduction of this loss, following Ref.~\cite{dung1998three} we introduce a bosonic noise operator $\phi_\alpha(\bx, \omega)$ for every position $\bx$ and for every frequency $\omega > 0$ which satisfy $[\phi_\alpha(\bx, \omega), \phi^\dagger_\alpha(\bx', \omega')] = \delta^3(\bx - \bx') \delta(\omega - \omega')$. Furthermore, the bath's Hamiltonian is expressible as\footnote{We use the notation that $\mathbb{R}^+ := [0, \infty)$},
\begin{equation}\label{eq:h_phi}
H_\text{bath} = \int_{\substack{\bx \in \mathbb{R}^3 \\ \omega \in \mathbb{R}^+ }} \omega \phi^\dagger_\alpha(\bx, \omega) \phi_\alpha(\bx, \omega) \ d^3 \bx\ d\omega.
\end{equation} 
We define the single-frequency field operator in the presence of loss, $\psi_\alpha(\bx, \omega)$, in terms of the noise operators $\phi_\alpha(\bx, \omega)$ by
\begin{align}\label{eq:fop_freq}
&\psi_\alpha(\bx, \omega) =\bigg(\frac{\omega^2 \alpha(\omega)}{\pi}\bigg)^{1/2} \times \nonumber\\
 &\qquad \int_{\bx' \in \mathbb{R}^3 } \text{G}_\alpha(\bx, \bx'; \omega) \phi_\alpha(\bx', \omega) d^3 \bx' d\omega,
\end{align}
where {\color{black}$\text{G}_\alpha(\bx, \bx'; \omega)$ is the Green's function corresponding to the lossy inhomogeneous bath, which is obtained by solving the partial differential equation
\begin{equation}
\big[\nabla_\bx^2  + \omega^2\varepsilon_\alpha(\bx, \omega)\big] \text{G}_\alpha(\bx, \bx'; \omega) = \delta^3(\bx - \bx').
\end{equation}
under the boundary condition that $\text{G}_\alpha(\bx, \bx'; \omega)$ is an outgoing wave as $|\bx|\to \infty$. We point out that since $\text{Im}[\varepsilon_\alpha(\bx, \omega)] = \alpha(\omega) > 0$, this boundary condition also implies that $\text{G}_\alpha(\bx, \bx';\omega)$ decays exponentially with $|\bx|$ as $|\bx|\to \infty$.} The field operators in the presence of loss, $\psi_\alpha(\bx)$ and $\pi_\alpha(\bx)$, are related to the single-frequency field operator via
\begin{subequations}
\begin{align} \label{eq:freq_decomp_ops}
&\psi_\alpha(\bx) = i\int_{\omega \in \mathbb{R}^+} \big(\psi_\alpha(\bx, \omega) - \psi_\alpha^\dagger(\bx, \omega)\big)d\omega, \\
&\pi_\alpha(\bx) = \varepsilon(\bx)\int_{\omega \in \mathbb{R}^+}\omega \big(\psi_\alpha(\bx, \omega) - \psi_\alpha^\dagger(\bx, \omega)\big) d\omega.
\end{align}
\end{subequations}
It is shown in appendix \ref{app:sec_quant_noise} that the single-frequency field operator $\psi_\alpha(\bx, \omega)$ has the commutator $[\psi_\alpha(\bx, \omega), \psi_\alpha^\dagger(\bx', \omega')] = - {\delta(\omega - \omega')} \text{Im}\big[\text{G}_\alpha(\bx, \bx'; \omega)\big] /{2\pi}$. Furthermore, this implies that the operators $\psi_\alpha(\bx)$ and $\pi_\alpha(\bx)$ satisfy the canonical commutation relations $[\psi_\alpha(\bx), \pi_\alpha(\by)] = i\delta^3(\bx - \by)$. Finally, we point out the field operators for a lossless bosonic bath, $\psi(\bx, \omega), \psi(\bx), \pi(\bx)$, can be constructed from the operators $\psi_\alpha(\bx, \omega), \psi_\alpha(\bx), \pi_\alpha(\bx)$ by taking $\alpha(\omega) \to 0$. Since $\lim_{\alpha(\omega) \to 0}\text{Im}\big[\text{G}_\alpha(\bx, \bx'; \omega)]$ exists and is well defined, this immediately implies the following commutator for the single-frequency field operator
\begin{align}\label{eq:comm_single_freq_op}
[\psi(\bx, \omega), \psi^\dagger(\bx', \omega')] = - \frac{\delta(\omega - \omega')}{2\pi} \text{Im}\big[\text{G}(\bx, \bx'; \omega)\big],
\end{align}
and the canonical commutation relation, $[\psi(\bx), \pi(\bx')] = i\delta^3(\bx - \bx')$, holds between the full field operators $\psi(\bx)$ and $\pi(\bx)$.

\subsubsection{Second quantization with plane-wave operators}
Since the physics of an inhomogeneous bath can fundamentally be captured by the fields scattered from the inhomogeneity when excited by incident plane-waves, we seek its second quantized description in terms of incident plane-wave annihilation operators. {\color{black}This can be derived from the noise-operator description by taking the limit of $\alpha(\omega) \to 0$. In order to take this limit, we express the Green's function $\text{G}_\alpha(\bx, \bx'; \omega)$ as
\begin{align}\label{eq:T_op_decomp_gfunc}
&\text{G}_\alpha(\bx, \bx'; \omega) = \nonumber \\
&\quad \text{G}_\alpha^0(\bx, \bx'; \omega) + \int_{\mathbb{R}^3} \text{S}_\alpha(\bx, \by; \omega) \text{G}_\alpha^0(\by, \bx'; \omega) d^3\by,
\end{align}
where $\text{G}_\alpha(\bx, \bx'; \omega)$ is the Green's function corresponding to the (lossy) background i.e.~it is the solution of the partial differential equation
\begin{align}
\big[\nabla_\bx^2  + \omega^2(\varepsilon_0 + i\alpha(\omega))\big] \text{G}_\alpha^0(\bx, \bx'; \omega) = \delta^3(\bx - \bx'),
\end{align}
with the boundary condition that $\text{G}_\alpha^0(\bx, \bx'; \omega)$ is an outgoing spherical wave when $|\bx| \to \infty$ and $\text{S}_\alpha(\bx, \by; \omega)$ is the position-domain representation of the classical scattering operator of the (lossy) inhomogeneous bath. }The scattering operator is the operator that maps an incident field to the corresponding fields scattered from the inhomogeneous permittivity distribution. Substituting this expansion into Eq.~\ref{eq:fop_freq}, we obtain
\begin{subequations}
\begin{equation} \label{eq:T_op_decomp_psi}
\psi_\alpha(\bx, \omega) = \psi^0_\alpha(\bx, \omega) + \int_{\mathbb{R}^3}\text{S}_\alpha(\bx, \by; \omega) \psi^0_\alpha(\by, \omega) d^3\by,
\end{equation}
where
\begin{equation}\label{eq:inc_fld_op}
\psi^0_\alpha(\bx, \omega) = \bigg(\frac{\omega^2 \alpha(\omega)}{\pi}\bigg)^{1/2}\int_{\mathbb{R}^3} \text{G}_\alpha^0(\bx, \bx'; \omega) \phi_\alpha(\bx'; \omega) d^3 \bx'.
\end{equation}
\end{subequations}
Physically, Eq.~\ref{eq:T_op_decomp_psi} expresses the fields inside the inhomogeneous bath as the sum of an incident field and a scattered field. Furthermore, since the operator $\psi^0(\bx; \omega)$ describes the excitations incident on the scattering region, it is expressible as a sum of plane-wave modes propagating in the background media. Formally, this is accomplished by introducing the plane-wave decomposition of $\text{G}_\alpha^0(\bx, \bx'; \omega)$,
\begin{equation}
\text{G}_\alpha^0(\bx, \bx'; \omega) = \frac{1}{(2\pi)^3} \int_{\substack{\Omega\in \mathbb{S}^2 \\ k \in \mathbb{R}^+}} \frac{e^{ik\Omega\cdot(\bx - \bx')}\ k^2 d^2\Omega\ dk}{\omega^2 \varepsilon_0 - k^2 + i\omega^2\alpha(\omega)},
\end{equation}
into Eq.~\ref{eq:inc_fld_op}. This allows us to express $\psi^0_\alpha(\bx; \omega)$ as
\begin{equation}\label{eq:inc_fld_op_a}
\psi^0_\alpha(\bx, \omega) = N_0(\omega) \int_{\Omega\in \mathbb{S}^2} a_\alpha(\bx, \Omega; \omega) e
^{ik_0(\omega) \Omega\cdot \bx}\ d^2\Omega,
\end{equation}
where $N_0(\omega) = ({\omega\sqrt{\varepsilon_0}}/{16\pi^3})^{1/2}$ and $k_0(\omega) = \omega \sqrt{\varepsilon_0}$ and
\begin{align}\label{eq:a_op}
&a_\alpha(\bx, \Omega, \omega) = \bigg(\frac{16\pi^2\omega \alpha(\omega)}{\sqrt{\varepsilon_0}}\bigg)^{1/2} e^{-ik_0(\omega)\Omega\cdot \bx}\times \nonumber\\
&\qquad \int_{\substack{\bx' \in \mathbb{R}^3 \\ {k \in \mathbb{R}^+}}}  \frac{e^{ik\Omega\cdot(\bx - \bx')}}{\omega^2 \varepsilon_0 - k^2 + i\omega^2 \alpha(\omega)} \phi_\alpha(\bx', \omega) \ d^3\bx' \ dk.
\end{align}
It is shown in appendix \ref{app:lim_a} that taking the limit of $\alpha(\omega) \to 0$ makes $a_\alpha(\bx, \Omega, \omega)$ independent of $\bx$, enabling us to define $a(\Omega, \omega) = \lim_{\alpha(\omega) \to 0} a_\alpha(\bx, \Omega, \omega)$. Furthermore, $a(\hOmega, \omega)$ satisfies the commutator
\begin{align}\label{eq:comm_pw_op}
[a(\Omega, \omega), a^\dagger(\Omega', \omega')] = \delta^2(\Omega - \Omega')\delta(\omega - \omega'),
\end{align}
and Eq.~\ref{eq:inc_fld_op_a} in the limit of $\alpha(\omega) \to 0$ reduces to
\begin{equation}
\psi^0(\bx, \omega) =N_0(\omega) \int_{\Omega \in \mathbb{S}^2} a(\Omega, \omega) e^{ik_0(\omega) \Omega\cdot \bx} d^2\Omega.
\end{equation}
The operators $a(\hOmega, \omega)$ can thus be interpreted as the annihilation operator for a plane-wave mode incident on the inhomogeneous bath from the surrounding background and the full field operator $\psi(\bx, \omega)$ is then given by the sum of the incident and scattered field operators
\begin{equation}
\psi(\bx, \omega) = N_0(\omega)\int_{\Omega\in \mathbb{S}^2} \mathcal{E}(\bx, \Omega; \omega) a(\Omega, \omega) d^2\Omega,
\end{equation}
where
\begin{subequations}\label{eq:far_field_profile}
\begin{align}
&\mathcal{E}(\bx, \Omega; \omega) = e^{ik_0(\omega) \Omega\cdot \bx} + \mathcal{E}_s(\bx, \hOmega; \omega) \text{ with } \\
 &\mathcal{E}_s(\bx, \hOmega; \omega) = \int_{\mathbb{R}^3} \text{S}(\bx, \by; \omega) e^{ik_0(\omega) \Omega\cdot \by}d^3\by,
\end{align}
\end{subequations}
with $\text{S}(\bx, \by; \omega) = \lim_{\alpha(\omega) \to 0} \text{S}_\alpha(\bx, \by; \omega)$. Finally, noting from Eqs.~\ref{eq:h_phi} and \ref{eq:a_op} that $[a(\Omega, \omega), H_\text{bath}] = \omega a(\Omega, \omega)$, and consequently
\begin{equation}
H_\text{bath} =\int_{\substack{\hOmega \in \mathbb{S}^2 \\ \omega \in \mathbb{R}^+}} \omega a^\dagger(\Omega, \omega) a(\Omega, \omega)  \ d^2\Omega\ d\omega.
\end{equation}
\section{Scattering from a localized two-level system}\label{sec:tls_scat}
In this section, we consider the problem of embedding a localized quantum system, which for simplicity is assumed to be a two-level system (TLS) with ground state $\ket{g}$ and excited state $\ket{e}$, in the inhomogeneous bath. We make the point-dipole approximation and assume that the localized quantum system interacts with the bath's field at only $\bx = \bx_d$. Furthermore, we assume a rotating-wave approximation for the interaction Hamiltonian for this two-level system to obtain
\begin{align}\label{eq:int_hamil}
H_\text{int} =iV_0\int_{\omega \in \mathbb{R}^+} \psi(\bx_d, \omega) \sigma^\dagger d\omega + \text{h.c.},
\end{align}
where $\sigma^\dagger = \ket{e}\bra{g}$, $\sigma = \ket{g}\bra{e}$ and $V_0$ is the coupling constant between the two-level system and the inhomogeneous bath. This Hamiltonian can be expressed in terms of the operators $a(\Omega, \omega)$ as
\begin{align}\label{eq:int_hamil}
&H_\text{int} = \nonumber \\
&iV_0 \int_{\substack{\hOmega \in \mathbb{S}^2 \\ \omega\in\mathbb{R}^+ }}N_0(\omega)\mathcal{E}(\bx_d, \Omega, \omega) a(\Omega, \omega) \sigma^\dagger \ d^2\Omega\ d\omega + \text{h.c.}
\end{align}
It can be pointed out that we do not make the Markovian approximation \cite{carmichael2009open, breuer2002theory} in this interaction Hamiltonian --- we intend our analysis to account for any strong frequency dependent interactions that can be induced by a spatial structuring of the bath. The total Hamiltonian $H$ for the system can be expressed as a sum of this interaction Hamiltonian, the Hamiltonian of the bath and the Hamiltonian of the two-level system: $H = H_0 + H_\text{int}$, where $H_0 = H_\text{bath} + \omega_0 \sigma^\dagger \sigma$.

{\color{black}We now consider the problem of exciting the two-level system with $N$ incident particles, and compute the spectral and spatial distribution of the scattered state. The input state that we consider is that of $N$ plane waves at frequencies $\vec{\nu} = (\nu_1, \nu_2 \dots \nu_N)$\footnote{Throughout this paper, we will adopt that notation that the symbol $\vec{f}$ represents the vector $(f_1, f_2 \dots f_N)$, where the length of the vector will be clear from the context.} and propagating in directions $\vec{\Omega} = (\Omega_1, \Omega_2 \dots \Omega_N)$
\begin{align}
\ket{\psi_\text{in}^N(\vec{\Omega}, \vec{\nu})} = \prod_{i=1}^N a^\dagger(\Omega_i, \nu_i) \ket{\text{vac}, g},
\end{align}
and we measure the overlap of the scattered state with $\ket{\psi_\text{out}(\vec{\bx}, \vec{\omega})}$, where
\begin{align}
\ket{\psi_\text{out}^N(\vec{\bx}, \vec{\omega})} = \prod_{i=1}^N \psi^\dagger(\bx_i, \omega_i) \ket{\text{vac}, g}.
\end{align}
This corresponds to monitoring the scattered $N-$particle quantum field at positions $\vec{\bx} = (\bx_1, \bx_2\dots \bx_N)$ and at frequencies $\vec{\omega} = (\omega_1, \omega_2 \dots \omega_N)$. The $N-$particle scattering matrix element corresponding to this measurement is thus given by
\begin{align}\label{eq:def_n_ph_smat}
&S^N\big(\vec{\bx}, \vec{\omega}; \vec{\Omega}, \vec{\nu}\big) = \bra{\psi_\text{out}(\vec{\bx}, \vec{\omega})} \hat{\text{S}}\ket{\psi_\text{in}(\vec{\Omega}, \vec{\nu})}
\end{align}
where $\hat{\text{S}}$ is the scattering matrix defined via
\begin{align}
\hat{\text{S}} = \lim_{\substack{t_f \to \infty \\ t_i \to -\infty}} e^{iH_0 t_f} e^{-i H(t_f - t_i)} e^{-i H_0 t_i}.
\end{align}}
Using this definition of $\hat{\text{S}}$, the scattering matrix-element can be expressed in terms of the Heisenberg picture field operators $\psi(\bx, \omega; t) = e^{iHt} \psi(\bx, \omega) e^{-iHt}$ and $a(\Omega, \omega; t) = e^{iH t} a(\Omega, \omega) e^{-iHt}$
\begin{align}\label{eq:smat_heis_pic}
&S^N\big(\vec{\bx}, \vec{\omega}; \vec{\Omega}, \vec{\nu}\big)= \lim_{\substack{t_f \to \infty \\ t_i \to -\infty}} e^{i\sum_{n=1}^N (\omega_n t_f - \nu_n t_i)} \times  \nonumber \\ &\qquad \langle g |\mathcal{T} \bigg[\prod_{n=1}^N \psi(\bx_n, \omega_n; t_f) \prod_{n = 1}^N a^\dagger(\Omega_n, \nu_n; t_i) \bigg]| g\rangle,
\end{align}
where $\mathcal{T}[\cdot]$ is the chronological time-ordering operator and we have used $e^{iH_0 t} \psi(\bx, \omega) e^{-iH_0 t} = \psi(\bx, \omega)e^{-i\omega t}$ and $e^{iH_0 t}a(\hOmega, \omega) e^{-iH_0 t} = a(\hOmega, \omega)e^{-i\omega t}$. Using the commutators in Eqs.~\ref{eq:comm_single_freq_op} and \ref{eq:comm_pw_op}, the Heisenberg equations of motion for $\psi(\bx, \omega; t)$ and $a(\hOmega, \omega; t)$ can be setup and integrated to obtain
\begin{subequations}\label{eq:heis_pic}
\begin{align}
&a(\Omega, \omega; t_f) = e^{-i\omega t_f} \bigg[a(\Omega, \omega; t_i) e^{i\omega t_i} - \nonumber\\
&\qquad \qquad V_0N_0(\omega) \mathcal{E}^*(\bx_d, \Omega; \omega) \int_{t_i}^{t_f} \sigma(\tau)e^{i\omega \tau} d\tau\bigg], \\
&\psi(\bx, \omega; t_f) = e^{-i\omega t_f}\bigg[\psi(\bx, \omega; t_i) e^{i\omega t_i} + \nonumber \\
&\qquad\qquad \frac{V_0}{\pi}\text{Im}\big[\text{G}(\bx, \bx_d; \omega)\big] \int_{t_i}^{t_f}\sigma(\tau) e^{i\omega \tau}d\tau\bigg],
\end{align}
\end{subequations}
where $\sigma(t) = e^{iHt} \sigma e^{-iHt}$. As is shown in appendix \ref{app:smat_to_gfunc}, Eqs.~\ref{eq:heis_pic} and \ref{eq:smat_heis_pic} can be used to obtain
\begin{align}\label{eq:smat_gen_exp}
&S^N(\vec{\bx}, \vec{\omega}; \vec{\Omega}, \vec{\nu}) =\bigg(\prod_{n=1}^N N_0(\nu_n)\bigg)\sum_{k=1}^N \bigg[\sum_{\mathcal{B}_k, \mathcal{D}_k} \bigg(\nonumber \sum_{\mathcal{P}_k}\prod_{n=1}^k\\
&\quad\mathcal{E}(\bx_{\mathcal{P}_k\mathcal{B}_k^N(n)}, \hOmega_{\mathcal{D}_k^N(n)}, \nu_{\mathcal{D}_k(n)}) \delta\big(\omega_{\mathcal{P}_k\mathcal{B}_k^N(n)} - \nu_{\mathcal{D}_k(n)}\big)\bigg) \times \nonumber \\ &\quad\bigg(\prod_{n=1}^{N-k}\frac{V_0^2}{\pi}\text{Im}\big[\text{G}(\bx_{\bar{\mathcal{B}}_k^N(n)}, \bx_d; \omega_{\bar{\mathcal{B}}_k^N(n)})\big] \times \nonumber \\
&\quad\mathcal{E}(\bx_d, \hOmega_{\bar{\mathcal{D}}_k^N(n)}, \nu_{\bar{\mathcal{D}}^N_k(n)})\bigg) \mathcal{G}^{N-k}\big(\vec{\omega}_{\bar{\mathcal{B}}^N_k} ;\vec{\nu}_ {\bar{\mathcal{D}}^N_k }\big)\bigg],
\end{align}
{\color{black}where the notation used in Eq.~\ref{eq:smat_gen_exp} is
\begin{enumerate}
\item $\mathcal{B}_k^N$ and $\mathcal{D}_k^N$ are two (independent) unordered $k-$element subsets of $\{1, 2 \dots N\}$,
\item $\bar{\mathcal{B}}_k^N$ and $\bar{\mathcal{D}}_k^N$ are complements of $\mathcal{B}_k^N$ and $\mathcal{D}_k^N$ respectively,
\item $\mathcal{P}_k$ is a $k-$element permutation,
\item $\mathcal{P}_k\mathcal{B}_k^N$ ($\mathcal{P}_k \mathcal{D}_k^N$) is a permutation of the $k-$element set $\mathcal{B}_k^N$ ($\mathcal{D}_k^N$),
\item For any set $\mathcal{A}$, $\mathcal{A}(i)$ is its $i^\text{th}$ element and
\item $\vec{\omega}_{\bar{B}_k^N} = (\omega_{\bar{\mathcal{B}}_k^N(1)}, \omega_{\bar{\mathcal{B}}_k^N(2)} \dots \omega_{\bar{\mathcal{B}}_k^N(N - k)} )$ (similarly for $\bar{D}_k^N \vec{\nu}$).
\end{enumerate}}
Furthermore, $\mathcal{G}^{N}(\vec{\omega}; \vec{\nu})$ is the $N-$excitation Green's function for the two-level system and is given by
\begin{align}\label{eq:nex_gfunc}
\mathcal{G}^N(\vec{\omega}; \vec{\nu}) = 
&\int_{\substack{t_1, t_2 \dots t_N \in \mathbb{R} \\ s_1, s_2 \dots s_N \in \mathbb{R}}}\bra{\text{vac}, g}\mathcal{T}\bigg[\prod_{i=1}^N \sigma(t_i)\sigma^\dagger(s_i) \bigg] \ket{\text{vac}, g}\times \nonumber \\
&\qquad\prod_{j=1}^N e^{i(\omega_j t_j - \nu_j s_j)}dt_j ds_j.
\end{align}
We note that the spatial dependence of the scattering matrix-element in Eq.~\ref{eq:smat_gen_exp} depends on $\mathcal{E}(\bx, \hOmega, \omega)$, which are the fields produced by the permittvity distribution on excitation by an incident plane-wave, as well $\text{Im}\big[\text{G}(\bx, \bx_d; \omega)]$ that captures the fields radiated by the two-level system. However, while $\text{Im}[\text{G}(\bx, \bx_d; \omega)]$ has contributions from both incoming and outgoing waves, when considering the action of the scattering-matrix on a normalizable $N-$particle wave-packet, only the outgoing waves have a non-zero contribution. Mathematically, this follows from the fact that $\text{G}^*(\bx, \bx'; \omega)$ is analytic in the lower-half of the complex plane and thus
\begin{align}
\lim_{t\to \infty} \int_0^\infty f(\omega) \text{G}^*(\bx, \bx'; \omega) e^{-i\omega t} d\omega = 0,
\end{align}
for any smooth and square-integrable function $f(\omega)$. Therefore, we can only retain the outgoing part of $\text{Im}\big[\text{G}(\bx, \bx_d; \omega)]$ in Eq.~\ref{eq:smat_gen_exp} and obtain
\begin{align}\label{eq:smat_gen_exp_2}
&S^N(\vec{\bx}, \vec{\omega}; \vec{\Omega}, \vec{\nu}) =\bigg(\prod_{n=1}^N N_0(\nu_n)\bigg)\sum_{k=1}^N \bigg[\sum_{\mathcal{B}_k, \mathcal{D}_k} \bigg(\nonumber \sum_{\mathcal{P}_k}\prod_{n=1}^k\\
&\quad\mathcal{E}(\bx_{\mathcal{P}_k\mathcal{B}_k^N(n)}, \hOmega_{\mathcal{D}_k^N(n)}, \nu_{\mathcal{D}_k(n)}) \delta\big(\omega_{\mathcal{P}_k\mathcal{B}_k^N(n)} - \nu_{\mathcal{D}_k(n)}\big)\bigg) \times \nonumber \\ &\quad\bigg(\prod_{n=1}^{N-k}\frac{V_0^2}{2\pi i}\text{G}(\bx_{\bar{\mathcal{B}}_k^N(n)}, \bx_d; \omega_{\bar{\mathcal{B}}_k^N(n)}) \times \nonumber \\
&\quad\mathcal{E}(\bx_d, \hOmega_{\bar{\mathcal{D}}_k^N(n)}, \nu_{\bar{\mathcal{D}}^N_k(n)})\bigg) \mathcal{G}^{N-k}\big(\vec{\omega}_{\bar{\mathcal{B}}^N_k} ;\vec{\nu}_ {\bar{\mathcal{D}}^N_k }\big)\bigg],
\end{align}
Equation \ref{eq:smat_gen_exp_2} is one of the key results of this paper --- to compute the quantum scattering matrix, we need to compute $\mathcal{E}(\bx, \Omega; \omega)$ and $\text{G}(\bx, \bx'; \omega)$, which can be obtained by solving the (classical) inhomogeneous wave equation, and $\mathcal{G}^N(\vec{\omega}; \vec{\nu})$, the Green's function  of the quantum emitter. {\color{black}Furthermore, the $N-$particle scattering matrix in Eq.~\ref{eq:smat_gen_exp_2}, consistent with the cluster decomposition principle\cite{weinberg1995quantum}, comprises of terms where $k < N$ particles are not scattered by the two-level system (which results in the $\delta(\cdot)$ functions conserving the individual frequencies) while the remaining $N-k$ particles interact with the quantum emitter (which is captured by the $N-k$ excitation Green's function $\mathcal{G}^{N-k}(\vec{\omega}; \vec{\nu})$).  }

In the following subsections, we explicitly calculate and study the properties of single- and two-particle scattering matrices. We provide analytical results for single and two-particle scattering when the two-level system is coupled to a homogeneous bath (i.e.~$\varepsilon(\bx) = \varepsilon_0$ for all $\bx$). We also numerically study properties of the single and two-particle scattering when the inhomogeneous bath is a 2D particleic crystal with a point defect. While we restrict our analysis to two-level emitters throughout this paper, in appendix \ref{app:mark_scat} we extend the formalism of this paper to an emitter with a general level structure but under the Markovian approximation.

\subsection{Single-particle scattering}
We first consider scattering of a single particle propagating in the inhomogeneous bath from the localized two-level system. The spectral properties of the single-particle scattering matrix are governed by the single-excitation Green's function
\begin{align}
&\mathcal{G}^1(\omega; \nu) = \nonumber \\ & \int_{t, s\in \mathbb{R}}
\bra{g, \text{vac}} \mathcal{T}\big[\sigma(t) \sigma^\dagger(s)\big]\ket{g, \text{vac}} e^{i(\omega t - \nu s)} dt ds.
\end{align}
Since $\sigma(t)\ket{\text{vac}, g} = 0$, it follows that $\bra{g, \text{vac}} \mathcal{T}\big[\sigma(t)\sigma^\dagger(s)\big] \ket{g, \text{vac}} = A_e(t-s) \Theta(t \geq s)$ where $A_e(t) = \bra{e, \text{vac}} e^{-iHt}\ket{e, \text{vac}}$. It then follows that
\begin{align}\label{eq:single_ex_gfunc_value}
\mathcal{G}^1(\omega; \nu) = 2\pi \delta(\omega - \nu)\underbrace{\int_0^\infty A_e(t)e^{i\nu t} dt}_{\mathcal{G}_0(\nu)}.
\end{align}
As is shown in appendix \ref{app:single_ex_gfunc}, $A_e(t)$ can be computed by solving the following (non-Markovian) ODE

\begin{align}\label{eq:single_ex_gfunc_value_ode}
&\frac{dA_e(t)}{dt} = -i\omega_0 A_e(t) + \frac{V_0^2}{\pi} \times\nonumber\\
&\qquad \int_{\substack{t' \in [0, t] \\ \omega \in \mathbb{R}^+ }} \text{Im}\big[\text{G}(\bx_d, \bx_d; \omega)\big] e^{-i\omega (t - t')} A_e(t') dt' \ d\omega.
\end{align}
subject to the boundary condition $A_e(t) = 0$ for $t \leq 0$. Specializing Eq.~\ref{eq:smat_gen_exp_2} to the case of single-particle scattering ($N = 1$), we obtain
\begin{align}
&S^1\big(\bx, \omega; \hOmega, \nu\big) = N_0(\nu)\bigg( e^{ik_0(\nu)\hOmega\cdot \bx} + \mathcal{E}_s(\bx, \hOmega, \nu) - \nonumber \\
&\qquad  iV_0^2  \mathcal{E}(\bx_d, \hOmega, \nu)\text{G}(\bx, \bx_d; \nu) \mathcal{G}_0(\nu)\bigg)\delta(\omega - \nu).
\end{align}
This form of the single-particle scattering matrix is immediately physically interpretable --- we notice that as a consequence of energy conservation, the frequencies of the incoming and outgoing single-particles are constrained to be equal. The spatial dependence of the scattering matrix can be seen to be the sum of three terms --- first is the incident plane-wave $e^{ik_0(\nu)\Omega\cdot \bx}$, second is the fields produced by direct scattering of the incident plane-wave from the inhomogeneous bath, $\mathcal{E}_s(\bx, \Omega, \nu)$, and third are dipolar fields radiated by the two-level system due to excitation by the incident plane-wave. We note that the strength of the dipolar field depends on the coupling strength $V_0$, as well as the total fields at the position of the dipole $\mathcal{E}(\bx_d, \hOmega, \nu)$.\\

\begin{figure*}[htpb]
\centering
\includegraphics[scale=0.23]{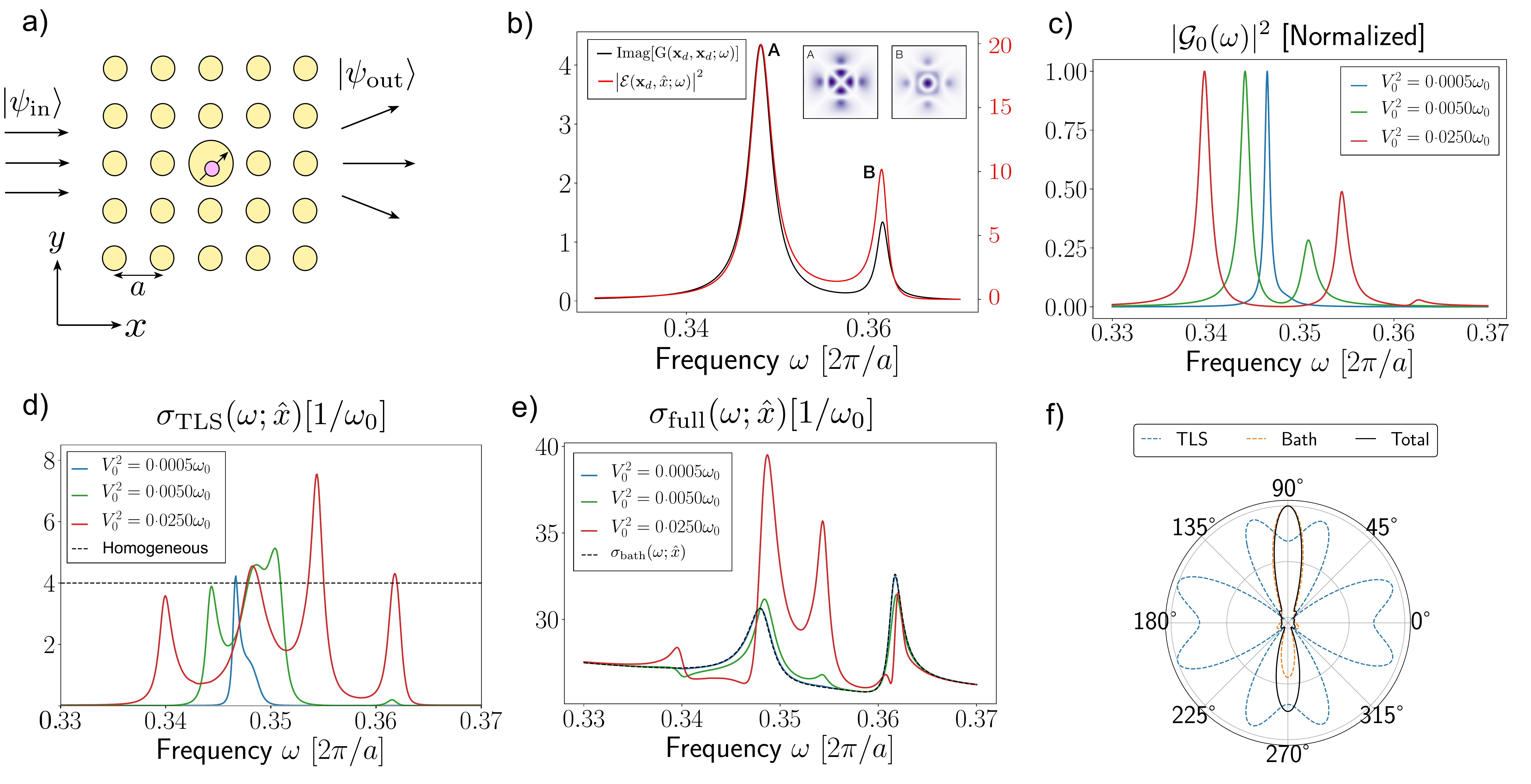}
\caption{\textbf{Single-particle scattering} from an inhomogeneous bosonic bath. a) Schematic depiction of the permittivity distribution --- a 2D square lattice photonic crystal cavity structure {\color{black}which is translationally invariant and infinite along the $z-$axis}. The radii of the central (defect) cylinder is $0.65a$, and of the surrounding cylinders is $0.2a$. The permittivity of the cylinder is assumed to be 8.9 and the two-level system is assumed to be positioned at $\bx_d = (0, 0.5a)$ with respect to the center of the central cylinder. b) Spectral dependence of the imaginary part of the Green's function $\text{Im}[G(\bx_d, \bx_d; \omega)]$ at the TLS position and the total electric field at the TLS position on excitation with a plane-wave propagating along the $x$ direction. The photonic crystal structure supports two (lossy) localized modes that are shown as an inset. c) The single-excitation Green's function (normalized to its maximum value) as a function of frequency for different coupling strengths $V_0$. The resonant frequency of the two-level system is assumed to be $\omega_0 = 0.35 (2\pi / a)$. d) The scattering cross-section corresponding to the fields emitted by the two-level system for different coupling strengths. e) The total scattering cross-section seen by the incident plane-wave field. f) Far field angular distribution of the fields scattered by the bath, the two-level system and the total scattered fields {\color{black}in the $xy$-plane as a function of the angle with respect to the $x-$axis}.}
\label{fig:single_particle_scat}
\end{figure*}
We first consider a homogeneous bath, for which $\mathcal{E}_s(\bx, \Omega, \nu) = 0$, and assume that $\bx_d = 0$ and $\mathcal{E}(\bx_d, \hOmega, \nu) = 1$. The spatial dependence of the single-particle scattering matrix is thus only determined by interference between the incident plane-wave and an outgoing spherical waves emitted by the two-level system at $\textbf{x} = \textbf{x}_d$
\begin{align}\label{eq:hom_3d_single_ph}
&S^1\big(\bx, \omega; \hOmega, \nu\big) = \nonumber\\
&\qquad N_0(\nu)\bigg( e^{ik_0(\nu)\hOmega\cdot \bx}  + iV_0^2  \frac{e^{ik_0(\nu)r}}{4\pi r} \mathcal{G}_0(\nu)\bigg)\delta(\omega - \nu).
\end{align}
Furthermore, the Weisskopf-Wigner theory \cite{wang1974wigner} can be used to approximate the single-excitation Green's function by a complex Lorentzian (refer to appendix \ref{app:single_ex_gfunc})
\begin{align}
\mathcal{G}_0(\nu) = \frac{1}{i(\omega_0 - \nu) + \gamma / 2},
\end{align}
where $\gamma = 2V_0^2 \text{G}(\bx = 0, \bx' = 0; \omega_0) = V_0^2\omega_0 \sqrt{\varepsilon_0}$. Our result for the single-particle scattering matrix Eq.~\ref{eq:hom_3d_single_ph} can be used to calculate the single-particle scattering cross-section by integrating the square of the amplitude of the scattered field over a spherical surface to obtain
\begin{align}
\sigma_\text{3D}(\nu) = \frac{V_0^4}{4\pi^2} |\mathcal{G}_0(\nu)|^2 = \frac{1}{\pi^2 \omega_0^2\varepsilon_0}\bigg[ \frac{\gamma^2 / 4}{(\nu - \omega_0)^2 + \gamma^2 / 4}\bigg].
\end{align}
This result agrees with the classical scattering cross-section for a dipole scatterer, and the single-particle scattering cross-section for a dipole two-level system coupled to a homogeneous bath derived in Ref.~\cite{liu2016quantum}. A similar analysis yields the following result for the scattering cross-section for a 2D problem
\begin{align}\label{eq:scat_fs}
\sigma_\text{2D}(\nu) = \frac{4}{\omega_0 \sqrt{\varepsilon_0}} \bigg[\frac{\gamma^2/4}{(\nu - \omega_0)^2 + \gamma^2/4}\bigg],
\end{align}
where $\gamma = V_0^2 / 2$. We note that at resonance ($\omega = \omega_0$), both $\sigma_\text{3D}(\nu)$ and $\sigma_\text{2D}(\nu)$ are independent of the coupling strength $V_0$, and are completely determined by the resonant frequency of the two-level system.

As an example of an inhomogeneous bath, we consider a 2D photonic crystal structure with a point defect (Fig.~\ref{fig:single_particle_scat}a). This structure supports two resonant modes that interact with the two-level system (Fig.~\ref{fig:single_particle_scat}b). The single-excitation Green's function of the two-level system interacting with the photonic crystal structure computed from Eqs.~\ref{eq:single_ex_gfunc_value} and \ref{eq:single_ex_gfunc_value_ode} is shown in Fig.~\ref{fig:single_particle_scat}c (refer to appendix \ref{app:single_ex_gfunc} for computational details). We observe that for low coupling strengths $V_0$, the single-excitation Green's function is approximately a Lorentzian in frequency and shows multiple resonant features at higher coupling strengths --- this is consistent with polaritonic splitting seen in strongly coupled cavity QED systems. To gain more insight into the modification of the scattering properties of the two-level system due to its interaction with the photonic crystal structure, we calculate the {\color{black}total scattering cross-section corresponding to fields scattered at frequency $\omega$ by the two-level system, $\sigma_\text{TLS}(\omega; \hOmega)$, the bath, $\sigma_\text{bath}(\omega; \hOmega)$, and both the two-level system and bath, $\sigma_\text{full}(\omega; \hOmega)$, when excited with an incident plane-wave propagating along $\Omega$}, 
\begin{subequations}
\begin{align}
&\sigma_\text{TLS}(\omega; \hOmega)=V_0^4 |\mathcal{E}(\bx_d, \hOmega; \omega)|^2 |\mathcal{G}_0(\omega)|^2 \times \nonumber \\
& \qquad \qquad \bigg[ \lim_{R\to\infty}\int_{\mathbb{S}^1} R\big |\text{G}(R\hOmega', \bx_d; \omega)\big |^2 d\hOmega' \bigg], \\
&\sigma_\text{bath}(\omega; \hOmega)= \lim_{R\to\infty}\int_{\mathbb{S}^1} R\big |\mathcal{E}_s(R\hOmega', \bx_d; \omega)\big |^2 d\hOmega', \\
&\sigma_\text{full}(\omega; \hOmega)=\lim_{R\to\infty}\int_{\mathbb{S}^1} R\bigg|\mathcal{E}_s(R\hOmega', \bx_d; \omega) - \nonumber\\
&\qquad \qquad iV_0^2 \mathcal{E}(\bx_d, \Omega; \omega) \mathcal{G}_0(\omega) \text{G}(R\hOmega', \bx_d; \omega)\bigg |^2 d\hOmega' .
\end{align}
\end{subequations}
Figures~\ref{fig:single_particle_scat}d shows {\color{black}$\sigma_\text{TLS}(\omega; \hat{x})$} as a function of frequency $\omega$ --- for small coupling strength, this scattering cross section resembles a single Lorentzian similar to the result in the homogeneous bath Eq.~\ref{eq:scat_fs} (the maximum homogeneous scattering cross-section, $4 / \omega_0 \sqrt{\varepsilon_0}$, is indicated with the black dashed line). For larger coupling strengths, this scattering cross-section is no longer a Lorentzian in frequency due to the strong frequency dependence of the inhomogeneous bath and can become significantly larger than the scattering cross-section in homogeneous bath. Figure \ref{fig:single_particle_scat}e shows $\sigma_\text{full}(\omega; \hat{x})$ as a function of frequency $\omega$ --- for small coupling strengths, this is dominated by scattering from the inhomogeneous bath (i.e.~$\sigma_\text{full}(\omega; \hat{x}) \approx \sigma_\text{bath}(\omega; \hat{x})$ which is shown in the dashed black line) while for larger coupling strengths, the contribution of the two-level system is observable. The angular distribution of the scattered single particle in the far field is shown in Fig.~\ref{fig:single_particle_scat}f.
\subsection{Two-particle scattering}
The two-particle spectrum of the scattered wave-packet is governed by the two-excitation Green's function $\mathcal{G}^2(\vec{\omega}; \vec{\nu})$, which is given by
\begin{align}\label{eq:two_ex_gfunc}
&\mathcal{G}^2(\vec{\omega}; \vec{\nu}) =\nonumber \\
&\int_{\substack{t_1, t_2 \in \mathbb{R} \\ s_1, s_2 \in \mathbb{R}}}\bra{g, \text{vac}} \mathcal{T}\big[\sigma(t_1)\sigma(t_2) \sigma^\dagger(s_1)\sigma^\dagger(s_2)\big]\ket{g, \text{vac}}\times \nonumber\\
&\qquad \qquad \prod_{i=1}^2 e^{i(\omega_i t_i - \nu_i s_i)} dt_i ds_i.
\end{align}
As is shown in Appendix \ref{app:two_ex_gfunc}, following the procedure introduced in Ref.~\cite{shi2015multiphoton}, this Green's function can be obtained entirely from the single-excitation Green's function
\begin{subequations}
\begin{align}
&\mathcal{G}^2(\vec{\omega}; \vec{\nu}) = \sum_{\mathcal{P}_2} \mathcal{G}^1(\omega_1; \nu_{\mathcal{P}_2(1)})\mathcal{G}^1(\omega_2; \nu_{\mathcal{P}_2(2)}) \nonumber\\
&\quad \underbrace{ -\pi \frac{\mathcal{G}_0(\nu_1)\mathcal{G}_0(\nu_2) \mathcal{G}_0(\omega_1) \mathcal{G}_0(\omega_2)}{\Gamma(\omega_1 + \omega_2)} \delta\bigg(\sum_{j=1}^2 \big(\omega_j - \nu_j\big)\bigg)}_{\mathcal{G}^2_\text{C}(\vec{\omega}; \vec{\nu})},
\end{align}
where $\mathcal{P}_2$ is a two element permutation and 
\begin{align}
\Gamma(E) = \int_{\mathbb{R}} \mathcal{G}_0(\omega) \mathcal{G}_0(E - \omega) \frac{d\omega}{2\pi}.
\end{align}
\end{subequations}
Specializing Eq.~\ref{eq:smat_gen_exp_2} provides the following expression for the two-particle scattering matrix
\begin{subequations}\label{eq:two_ph_smat}
\begin{align}
&S^2\big(\vec{\bx}, \vec{\omega}; \vec{\Omega}, \vec{\nu}\big) = \sum_{\mathcal{P}_2} \bigg(S\big(\bx_1, \omega_1; \hOmega_{\mathcal{P}(1)}, \nu_{\mathcal{P}_2(1)}\big) \times \nonumber \\
&\qquad S\big(\bx_2, \omega_2; \hOmega_{\mathcal{P}(2)}, \nu_{\mathcal{P}_2(2)}\big)\bigg) +  S_\text{C}^2\big(\vec{\bx}, \vec{\omega}; \vec{\Omega}, \vec{\nu}\big),
\end{align}
where $\mathcal{P}$ is a permutation of the set $\{1, 2\}$, $S^1\big(\bx, \omega; \hOmega, \nu)$ is the single-particle scattering matrix and $S_\text{C}^2(\vec{\bx}, \vec{\omega}; \vec{\hOmega}, \vec{\nu})$ is the connected part of the two-particle scattering matrix that is given by
\begin{align}\label{eq:two_ph_conn_smat}
&S_\text{C}^2\big(\vec{\bx}, \vec{\omega}; \vec{\Omega}, \vec{\nu}\big) = -\frac{V_0^4}{4\pi^2} \bigg( \prod_{i=1}^2N_0(\nu_i) \mathcal{E}(\bx_d, \hOmega_i, \nu_i) \times \nonumber\\
&\qquad\qquad \text{G}(\bx_i, \bx_d; \omega_i)\bigg)\mathcal{G}_\text{C}^2(\vec{\omega}; \vec{\nu}).
\end{align}
\end{subequations}
We point out that Eq.~\ref{eq:two_ph_smat} is consistent with the form of the two-particle scattering matrix as expected by the cluster-decomposition principle \cite{weinberg1995quantum}, with the connected part of the scattering matrix capturing the particle-particle interactions induced by the nonlinearity of the two-level system. As can be seen from Eq.~\ref{eq:two_ph_conn_smat}, the spectral properties of the entanglement generated in the scattered two-particles due to such particle-particle interactions is captured by the connected part of the two-excitation Green's function, while the spatial distribution of the entanglement is simply given by the Green's function corresponding to the permittivity distribution.

For a homogeneous bath ($\varepsilon(\bx) = \varepsilon_0$) and assuming $\bx_d = 0$, the connected part of the scattering matrix reduces to a product of two outgoing radially symmetric spherical waves
\begin{align}
S_\text{C}^2\big(\vec{\bx}, \vec{\omega}; \vec{\Omega}, \vec{\nu}\big) = -\frac{V_0^4}{4\pi^2}\bigg(\prod_{i=1}^2 N_0(\nu_i) \frac{e^{ik_0(\omega_i)r_i}}{4\pi r_i}\bigg) \mathcal{G}_\text{C}^2(\vec{\omega}; \vec{\nu}),
\end{align}
where, in the  Weisskopf-Wigner approximation, $\mathcal{G}_\text{C}(\vec{\omega}; \vec{\nu})$ is given by (refer to appendix \ref{app:two_ex_gfunc} for details)
\begin{align}\label{eq:gfunc_conn_part}
\mathcal{G}_\text{C}^2(\vec{\omega}; \vec{\nu}) = -\frac{i\pi(\sum_{j=1}^2 \omega_j  - 2\omega_0 - i\gamma) \delta(\sum_{j=1}^2 (\omega_j - \nu_j))}{\prod_{j=1}^2(\nu_j - \omega_0 - i\gamma / 2)(\omega_j - \omega_0 - i\gamma/2)},
\end{align}
where $\gamma =  V_0^2\omega_0 \sqrt{\varepsilon_0}$. We point out that this result for the connected part of the two-excitation Green's function is identical to the one obtained in Waveguide QED under a Markovian approximation to the waveguide two-level system interaction \cite{fan2010input}.
\begin{figure*}[t]
\centering
\includegraphics[scale=0.265]{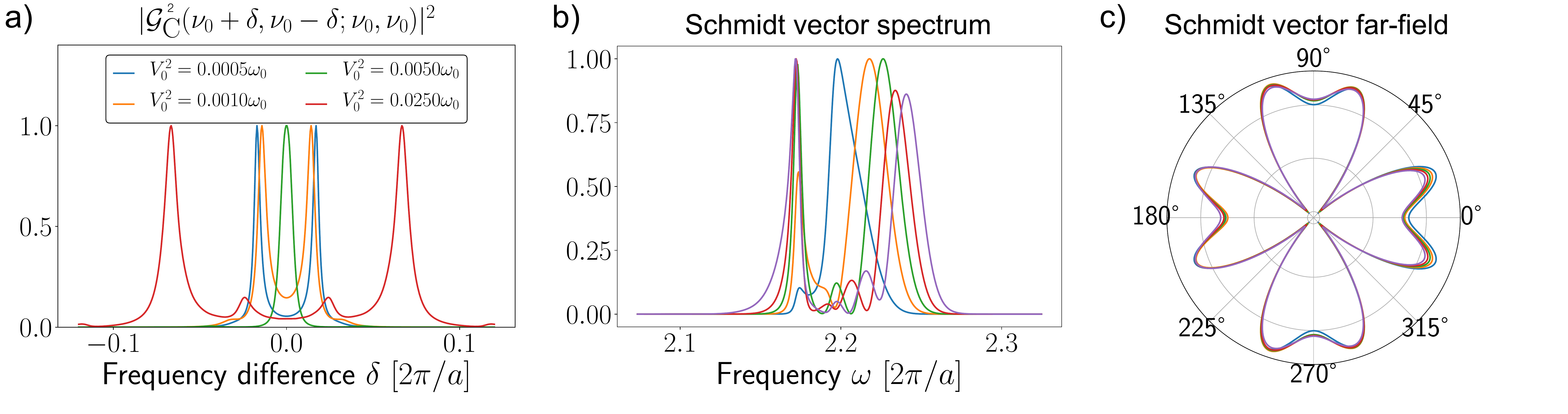}
\caption{\textbf{Two-particle scattering} from a localized two-level system embedded in the inhomogeneous bath shown in Fig.~\ref{fig:single_particle_scat}a. We assume both the incident particles to be propagating along the $x$ axis (i.e.~$\Omega_1 = \Omega_2 = \hat{x}$) for our simulations. a) The connected part of the two-excitation Green's function as a function of the frequency difference between the two scattered particles assuming that the incident particles are at frequency $\nu_0$ resonant with the two-level system ($\nu_0 = \omega_0$). b) The frequency spectrum and c) the far field angular distribution of the first 5 Schmidt vectors of the output state corresponding to the connected part of the two-particle scattering matrix when both the input particles are Gaussian wave-packets in frequency (Eq.~\ref{eq:inp_state}) with $\tau = 220 / \omega_0$.}
\label{fig:two_particle_scat}
\end{figure*}

Next, we consider two-particle scattering from the two-level system when it is embedded in the photonic crystal structure shown in Fig.~\ref{fig:single_particle_scat}a. Figure \ref{fig:two_particle_scat}b shows the dependence of the connected part of the two-excitation Green's function on the frequency difference of the output particles computed using Eq.~\ref{eq:gfunc_conn_part}. To study the spectral and spatial entanglement properties of the outgoing two-particles, we consider exciting the two-level system with a two-particle Fock state with both particles propagating along the $\hat{{x}}$ axis but with a Gaussian spectrum:
\begin{align}\label{eq:inp_state}
\ket{\psi_\text{in}} \propto \bigg( \int_{\mathbb{R}^+} e^{-(\nu - \omega_0)^2 \tau^2 / 2} a^\dagger(\hat{x}, \nu) d\nu\bigg)^2 \ket{\text{vac}, g},
\end{align}
where $\tau$ governs the spectral bandwidth of the incident wave-packet. For large $\omega_0 \tau \gg 1$, application of the connected part of the two-particle scattering matrix on this wave-packet yields the following two-particle wave-packet $\psi_\text{out,C}(\vec{\bx}, \vec{\omega})$
\begin{align}
&\psi_{\text{out,C}}(\vec{\bx}, \vec{\omega}) \approx -\frac{V_0^4 N_0(\omega_0) \mathcal{E}^2(\bx_d, \hat{x}, \omega_0)\mathcal{G}_0^2(\omega_0)}{4\pi^2} \times \nonumber\\
&\qquad \bigg(\prod_{i=1}^2 \mathcal{G}_0(\omega_i)\text{G}(\bx_i, \bx_d; \omega_i)\bigg) \frac{e^{-(\omega_1 + \omega_2 - 2\omega_0)^2\tau^2}}{\Gamma(\omega_1 + \omega_2)}.
\end{align}
The Schmidt decomposition of $\psi_{\text{out,C}}(\vec{\bx}, \vec{\omega})$ can then express it as a superposition of two-particle fock states
\begin{align}
\psi_{\text{out,C}}(\vec{\bx}, \vec{\omega}) = \sum_{k=1}^\infty \lambda_i \psi_i(\bx_1, \omega_1) \psi_i(\bx_2, \omega_2),
\end{align}
with $\lambda_1 \geq \lambda_2 \geq \lambda_3 \dots $. Figures \ref{fig:two_particle_scat}b and c show the spectral and spatial dependence of the first five Schmidt components of $\psi_{\text{out,C}}$ for a two-level system strongly coupled to the photonic crystal bath. The frequency spectrum of each Schmidt vector (shown in Fig.~\ref{fig:two_particle_scat}b) is computed by integrating the magnitude square of the Schmidt vector over a sufficiently large circle surrounding the photonic crystal structure and the angular distribution of the far field (shown in Fig.~\ref{fig:two_particle_scat}c) is computed by integrating the magnitude square of the Schmidt vector with respect to frequency. The multi-Lorentzian nature of the spectrum of the emitted particles can be attributed to a strong hybridization between the two-level system and the electromagnetic modes of the photonic crystal cavity structure, while the far field pattern is almost completely determined by the field profile of the photonic crystal cavity mode.

\section{Conclusion}
In conclusion, we presented an approach to exactly analyze spatial and spectral properties of few-particle scattering from a localized system embedded in a an inhomogeneous bath. We develop exact solutions for the single and two-particle scattering matrices when the localized system is a two-level system. The formalism presented in this paper paves the way for computationally understanding the impact of structured media on quantum scattering from embedded localized systems. While the formalism in this paper assumes that the bosonic bath can be described by an inhomogeneous scalar wave equation, it can  be extended to quantum optical systems where the bosonic bath (i.e.~the photonic field) interacting with the localized system is described by a vector wave-equation.
\section{Acknowledgements}
We thank David A.~B.~Miller and Geun Ho Ahn for providing feedback on the manuscript. R.~T.~acknowledges support from Kailath Graduate fellowship. This work is funded by the Air Force Office of Scientific Research under AFOSR MURI programme (award no.~FA9550-17-1-0002).

\bibliographystyle{IEEEtran}
\bibliography{library_v2.bib}

\appendix
\onecolumngrid
\section{Second quantization with noise operators}\label{app:sec_quant_noise}
We begin by outlining some useful properties of the scalar Green's function corresponding to the inhomogeneous wave-equation, followed by showing that the noise-operator based quantum theory presented in the main text satisfies the canonical commutation relations stemming from Dirac's quantization.
\subsection{Properties of the scalar inhomogeneous Green's function}
\noindent Throughout this section, $\text{G}_\alpha(\bx, \bx'; \omega)$ will refer to the scalar's green's function which satisfies
\begin{align}\label{eq:gfunc_def}
\big[\nabla^2_\bx + \omega^2\varepsilon_\alpha(\bx, \omega)\big]\text{G}_\alpha(\bx, \bx';\omega) = \delta^3(\bx - \bx'),
\end{align}
where $\varepsilon_\alpha(\omega) = \varepsilon(\bx) + i\alpha(\omega)$. Here $\varepsilon(\bx)$ is the inhomogeneous permittivity distribution corresponding to the inhomogenuous media and $\alpha(\omega)$ governs the loss in the media. {\color{black}We point out that while physically $\omega > 0$, it is convenient to consider the solution of Eq.~\ref{eq:gfunc_def} for $\omega < 0$ as well. We assume that for $\omega > 0, \alpha(-\omega) = -\alpha(\omega)$ --- this ensures that for $\omega < 0, \alpha(\omega) < 0$ and hence the outgoing solution to Eq.~\ref{eq:gfunc_def} for $\omega < 0$ decays exponentially as $|\bx| \to \infty$. Furthermore, we also make a physically reasonable assumption of $\alpha(\omega) \to 0$ as $|\omega| \to \infty$.}\\ \ \\ 
\noindent\textbf{Property 1}:
\begin{align}\label{eq:gfunc_prop_1}
\text{G}_\alpha(\bx, \bx'; \omega) = \text{G}_\alpha^*(\bx, \bx'; -\omega).
\end{align}
\emph{Proof}: This immediately follows by conjugating Eq.~\ref{eq:gfunc_def} and by noting that $\varepsilon_\alpha(\bx, -\omega) = \varepsilon_\alpha^*(\bx, \omega)$. \\ \ \\ \ \\
\noindent \textbf{Property 2}:
\begin{align}\label{eq:gfunc_prop_2}
2i \omega^2 \alpha  \int_{\mathbb{R}^3}\text{G}_\alpha^*(\bx', \by; \omega) \text{G}_\alpha(\bx, \by; \omega) d^3 \by = \text{G}_\alpha^*(\bx, \bx'; \omega) - \text{G}_\alpha(\bx, \bx'; \omega).
\end{align}
\noindent \emph{Proof}: We note that the scalar Green's function is necessarily reciprocal i.e.~$\text{G}_\alpha(\bx, \by; \omega) = \text{G}_\alpha(\by, \bx; \omega)$. Therefore, it follows from Eq.~\ref{eq:gfunc_def} that
\begin{align}
\big[\nabla^2_\by + \omega^2 \varepsilon_\alpha(\by; \omega)\big] \text{G}_\alpha(\bx, \by; \omega) = \delta^3(\bx - \by).
\end{align}
Multiplying this equation with $\text{G}_\alpha^*(\bx', \by; \omega)$ and integrating with respect to $\by$ we obtain
\begin{align}\label{eq:mult_and_int}
&\text{G}_\alpha^*(\bx, \bx'; \omega) = -\int_{\mathbb{R}^3} \nabla_\by \text{G}_\alpha^*(\bx', \by; \omega) \cdot \nabla_\by \text{G}_\alpha(\bx, \by; \omega) d^3 \by + \omega^2\int_{\mathbb{R}^3} \varepsilon_\alpha(\by; \omega) \text{G}_\alpha^*(\bx', \by; \omega) \text{G}_\alpha(\bx, \by; \omega) d^3 \by.
\end{align}
Conjugating Eq.~\ref{eq:mult_and_int} followed by swapping $\bx$ and $\bx'$, we obtain
\begin{align}\label{eq:mult_and_int_2}
&\text{G}_\alpha(\bx, \bx'; \omega) = -\int_{\mathbb{R}^3} \nabla_\by \text{G}_\alpha^*(\bx', \by; \omega) \cdot \nabla_\by \text{G}_\alpha(\bx, \by; \omega) d^3 \by + \omega^2\int_{\mathbb{R}^3} \varepsilon_\alpha^*(\by; \omega) \text{G}_\alpha^*(\bx', \by; \omega) \text{G}_\alpha(\bx, \by; \omega) d^3 \by.
\end{align}
Subtracting Eq.~\ref{eq:mult_and_int_2} from Eq.~\ref{eq:mult_and_int}, we obtain Eq.~\ref{eq:gfunc_prop_2}. \\ \ \\ \ \\
\noindent \textbf{Property 3}: Assuming that $\exists\ \varepsilon_\text{ub}$ such that $ \varepsilon(\bx) < \varepsilon_\text{UB} \ \forall \ \bx$ and $\varepsilon(\bx) > 0 \ \forall \ \bx$,
\begin{align}\label{eq:gfunc_prop_3}
\int_{-\infty}^\infty \omega \text{G}_\alpha(\bx, \bx'; \omega)\ d\omega  = -\frac{i\pi}{\varepsilon(\bx)}\delta^3(\bx - \bx').
\end{align}
\emph{Proof}: Defining $\theta(\bx) = \varepsilon_\text{ub} - \varepsilon(\bx)$, Eq.~\ref{eq:gfunc_def} can be rewritten as
\begin{align}\label{eq:gfunc_def_rearr}
\big[\nabla^2_\bx + \omega^2 (\varepsilon_\text{ub} + i\alpha(\omega))\big]\text{G}_\alpha(\bx, \bx'; \omega) = \omega^2 \theta(\bx) \text{G}_\alpha(\bx, \bx'; \omega) + \delta^3(\bx - \bx').
\end{align}
A Dyson series for $\text{G}_\alpha(\bx, \bx'; \omega)$ can be deduced from Eq.~\ref{eq:gfunc_def_rearr} to obtain
\begin{align}\label{eq:dyson}
&\text{G}_\alpha(\bx, \bx'; \omega)  = \text{G}_\alpha^\text{ub}(\bx, \bx'; \omega) + \sum_{n=1}^\infty \omega^{2k} \int_{\by_1, \by_2 \dots \by_k \in \mathbb{R}^3}  \bigg[\prod_{k=0}^{n}\text{G}_\alpha^\text{ub}(\by_k, \by_{k + 1}; \omega)\bigg]_{\substack{\by_0 = \bx \\ \by_{n + 1} = \bx'}} \bigg[\prod_{k=1}^n \theta(\by_k) d^3\by_k\bigg] ,
\end{align}
where $\text{G}_\alpha^\text{ub}(\bx, \bx'; \omega)$ satisfies
\begin{align}\label{eq:gfunc_eps_M_def}
\big[\nabla_\bx^2 + \omega^2(\varepsilon_\text{ub} + i\alpha(\omega))\big] \text{G}_\alpha^\text{ub}(\bx, \bx'; \omega) = \delta^3(\bx - \bx').
\end{align}
It can easily be seen from Eq.~\ref{eq:gfunc_eps_M_def} that
\begin{align}
\text{G}_\alpha^\text{ub}(\bx, \bx'; \omega) = \int_{\mathbb{R}^3} \frac{e^{i\bk \cdot (\bx - \bx')}}{\omega^2 \varepsilon_\text{ub} - |\bk|^2 + i\omega^2 \alpha(\omega)}\frac{d^3 \bk}{(2\pi)^3}.
\end{align}
 We next evaluate the integral in Eq.~\ref{eq:gfunc_prop_3} using the expansion for $\text{G}_\alpha(\bx, \bx'; \omega)$ in Eq.~\ref{eq:dyson}. Consider the following integral
 \begin{align}
 \int_{-\infty}^\infty \omega^{2k + 1} &\bigg[\prod_{k=0}^n \text{G}_\alpha^\text{ub}(\by_k, \by_{k + 1}; \omega)\bigg] d\omega = \int_{\omega \in \mathbb{R}} \int_{\bk_0, \bk_1 \dots \bk_{n}\in \mathbb{R}^3} \omega^{2k + 1}\bigg[\prod_{k=0}^n \frac{e^{i\bk_k \cdot (\by_k - \by_{k + 1})}}{\omega^2 \varepsilon_\text{ub} - |\bk_k|^2 + i\omega^2 \alpha(\omega)} \frac{d^3 \bk_k}{(2\pi)^3}\bigg]d\omega.
 \end{align}
 Noting that the integrand is analytic as a function of $\omega$ in the upper complex half-plane, it follows that the integral with respect to $\omega$ can be evaluated along $\omega = Re^{i\theta}$, with $R \to \infty$ with $\theta$ going from $\pi$ to $0$. This immediately yields
 \begin{align}
  \int_{-\infty}^\infty \omega^{2k + 1} \bigg[\prod_{k=0}^n \text{G}_\alpha^\text{ub}(\by_k, \by_{k + 1}; \omega)\bigg] d\omega = -\frac{i\pi}{\varepsilon_\text{ub}} \prod_{k=0}^n \delta^3(\by_k - \by_{k + 1}).
 \end{align}
 It thus follows from Eq.~\ref{eq:dyson} that
 \begin{align}
 \int_{-\infty}^\infty \omega \text{G}_\alpha^{\text{ub}}(\bx, \bx'; \omega) = -i\pi \delta^3(\bx - \bx')\sum_{k=0}^\infty \frac{\theta^k(\bx)}{\varepsilon_\text{ub}^{k + 1}} = -\frac{i \pi}{\varepsilon(\bx)}\delta^3(\bx - \bx').
 \end{align}
 \subsection{Verifying canonical commutation relations between $\psi(\bx)$ and $\pi(\bx)$}
\noindent We begin by deriving the commutator $[\psi(\bx, \omega), \psi^\dagger(\bx', \omega')]$.  Using Eq.~\ref{eq:fop_freq} and the commutator for the noise operator $\psi(\bx, \omega)$, we obtain
\begin{align}
[\psi(\bx, \omega), \psi^\dagger(\bx'; \omega')] = \frac{1}{\pi}\omega^2 \alpha(\omega)\delta(\omega - \omega')\int_{\mathbb{R}^3}\text{G}_\alpha^*(\bx', \by; \omega) \text{G}_\alpha(\bx, \by; \omega) d^3 \by.
\end{align}
Using Eq.~\ref{eq:gfunc_prop_2}, we obtain
\begin{align}\label{eq:freq_dom_comm}
[\psi(\bx, \omega), \psi^\dagger(\bx'; \omega')] &=\frac{\delta(\omega - \omega')}{2\pi i}  \big[ \text{G}_\alpha^*(\bx, \bx'; \omega) - \text{G}_\alpha(\bx, \bx'; \omega)\big],\nonumber\\ &= -\frac{\delta(\omega - \omega')}{2\pi}\text{Im}\big[\text{G}_\alpha(\bx, \bx'; \omega)\big].
\end{align}
Consider now the commutator $[\psi(\bx), \pi(\bx')]$: Using Eqs.~\ref{eq:freq_decomp_ops} and \ref{eq:freq_dom_comm}, we obtain
\begin{align}
[\psi(\bx), \pi(\bx')] = \frac{\varepsilon(\bx')}{\pi} \int_{\omega \in \mathbb{R}^+} \omega \big[ \text{G}_\alpha^*(\bx, \bx'; \omega) - \text{G}_\alpha(\bx, \bx'; \omega)\big] d\omega.
\end{align}
Using Eqs.~\ref{eq:gfunc_prop_1} and \ref{eq:gfunc_prop_3}, it immediately follows that $[\psi(\bx), \pi(\bx')] = i\delta^3(\bx - \bx')$ --- the noise operator based quantization thus reproduces the canonical commutation relations for inhomogeneous bosonic baths.

\section{Evaluating $a_\alpha(\bx, \hOmega, \omega)$ as $\alpha(\omega) \to 0$}\label{app:lim_a}
\noindent In this appendix, we detail the evaluation of the $\lim_{\alpha(\omega) \to 0} a_\alpha(\bx, \Omega; \omega)$, where $a_\alpha(\bx, \Omega; \omega)$ is defined in Eq.~\ref{eq:a_op}. In particular, we show that in the limit of $\alpha(\omega) \to 0$, $a_\alpha(\bx, \Omega; \omega)$ becomes independent of $\bx$. Since $a_\alpha(\bx, \Omega; \omega)$ are, by construction, bosonic annihilation operators, in order to show that $\lim_{\alpha(\omega) \to 0} a_\alpha(\bx, \Omega; \omega)$ is independent of $\bx$, it is sufficient to show that $\lim_{\alpha(\omega) \to 0} [a_\alpha(\bx, \Omega; \omega), a_\alpha^\dagger(\bx', \Omega'; \omega')]$ is independent of $\bx$ and $\bx'$. Using Eq.~\ref{eq:a_op} and $[\phi_\alpha(\bx; \omega), \phi^\dagger_\alpha(\bx'; \omega')] = \delta^3(\bx - \bx') \delta(\omega - \omega')$, we can explicitly evaluate this commutator to obtain
\begin{align}
[a_\alpha(\bx,& \Omega; \omega), a_\alpha^\dagger(\bx', \Omega'; \omega')] =\nonumber \\ &\frac{\alpha(\omega) \omega}{4\pi^4 \sqrt{\varepsilon_0}}\delta(\omega - \omega')\int_{\substack{\by \in \mathbb{R}^3 \\ k, k' \in \mathbb{R}^+}} \frac{k^2 k'^2 e^{i(k\Omega - k'\Omega')\cdot \by} e^{i(k - k_0(\omega))\Omega\cdot \bx} e^{-i(k' - k_0(\omega))\Omega'\cdot \bx'}}{(\omega^2 \varepsilon_0 - k^2 + i\omega^2 \alpha(\omega))(\omega^2 \varepsilon_0 - k'^2 - i\omega^2 \alpha(\omega))} dk dk' d^3\by.
\end{align}
Noting that
\begin{align}
\int_{\mathbb{R}^3} e^{i(k\Omega - k'\Omega')\cdot \by} d^3 \by =  \frac{8\pi^3\delta(k - k') \delta^2(\Omega - \Omega')}{k^2},
\end{align}
we obtain
\begin{align}
[a_\alpha(\bx, \Omega; \omega), a_\alpha^\dagger(\bx', \Omega'; \omega')] = \frac{2\alpha(\omega) \omega}{\pi \sqrt{\varepsilon_0}}\delta(\omega - \omega')\delta^2(\Omega - \Omega') \int_{\mathbb{R}^+} \frac{k^2 e^{i(k - k_0(\omega)) \Omega\cdot(\bx - \bx')}}{(\omega^2 \varepsilon_0 - k^2)^2 + \omega^4 \sigma^2(\omega)} dk.
\end{align}
We note that in the limit of $\alpha(\omega) \to 0$, the integral in the above equation can be approximated as follows
\begin{align}
\int_{\mathbb{R}^+} \frac{k^2 e^{i(k -  k_0(\omega)) \Omega\cdot(\bx - \bx')}}{(\omega^2 \varepsilon_0 - k^2)^2 + \omega^2 \sigma^4(\omega)} dk \approx \frac{1}{4} \int_{\mathbb{R}^+} \frac{1}{(k - \omega \sqrt{\varepsilon_0})^2 + \omega^2 \sigma^2(\omega) / 4\varepsilon_0} \approx \frac{\pi \sqrt{\varepsilon_0}}{2 \omega \alpha(\omega)}.
\end{align}
Consequently, we obtain
\begin{align}
\lim_{\alpha(\omega) \to 0} [a_\alpha(\bx, \Omega; \omega), a_\alpha^\dagger(\bx', \Omega'; \omega')] = \delta(\omega - \omega') \delta^2(\Omega - \Omega').
\end{align}
Thus, we conclude that the operators $a_\sigma(\bx, \Omega; \omega)$ are independent of $\bx$ in the limit of $\alpha(\omega) \to 0$. Defining $a(\Omega; \omega)$ via $a(\Omega; \omega) = \lim_{\alpha(\omega) \to 0} a_\alpha(\bx, \Omega; \omega)$, we thus obtain the following commutator
\begin{align}
[a(\Omega; \omega), a^\dagger(\Omega'; \omega')] =  \delta(\omega - \omega')\delta^2(\Omega - \Omega').
\end{align}

\section{Relating scattering matrix to the localized system's Green's functions}\label{app:smat_to_gfunc}
Our starting point is the Heisenberg picture representation of the $N-$particle scattering matrix element in Eq.~\ref{eq:smat_heis_pic}. We first substitute for $\psi(\bx, \omega; t_f)$ from Eq.~\ref{eq:heis_pic} to obtain
\begin{align}\label{eq:simp_smat}
&e^{i\sum_{n=1}^N \omega_n t_f}\bra{\text{vac}, g} \mathcal{T}\bigg[\prod_{i=1}^N \psi(\bx_i, \omega_i; t_f) \prod_{i=1}^N a^\dagger(\hOmega_i,\nu_i; t_i)\bigg]\ket{\text{vac}, g} \nonumber\\
&=\bra{\text{vac}, g} \sum_{k=0}^N \sum_{\mathcal{B}_k^N} \mathcal{T}\bigg[\bigg(\prod_{n\in \bar{\mathcal{B}}_k^{N}} \frac{V_0}{\pi}\text{Im}\big[\text{G}(\bx_n, \bx_d; \omega_n)\big] \int_{t_i}^{t_f}\sigma(\tau)e^{i\omega_n \tau}d\tau\bigg) \bigg( \prod_{n \in \mathcal{B}_k^N} \psi(\bx_n, \omega_n; t_i)e^{i\omega_n t_i}\bigg)\times \nonumber\\
&\bigg(\prod_{n=1}^N a^\dagger(\hOmega_n, \nu_n; t_i)\bigg) \bigg] \ket{\text{vac}, g},
\end{align}
where $\mathcal{B}_k^N$ is an unordered $k-$element subset of $\{1, 2, \dots N\}$, and $\bar{\mathcal{B}}_k^N$ is its complement. Using the commutator $[\psi(\bx, \omega), a^\dagger(\hOmega, \nu)] = N_0(\nu)\mathcal{E}(\bx, \hOmega, \nu) \delta(\omega - \nu)$ and the relation $\psi(\bx, \omega; t_i)\ket{\text{vac}, g} = 0$, we obtain
\begin{align}\label{eq:app_comm}
&\bigg(\prod_{n\in \mathcal{B}_k^N} \psi(\bx_n, \omega_n; t_i) \bigg) \prod_{n=1}^N a^\dagger(\hOmega_n, \nu_n; t_i) \ket{\text{vac}, g} =\nonumber \\
&\sum_{\mathcal{D}_k^N, \mathcal{P}_k} \bigg( \prod_{n = 1}^{N - k} a^\dagger(\hOmega_{\bar{\mathcal{D}}_k^N(n)}, \nu_{\bar{\mathcal{D}}_k^N(n)}; t_i)  \bigg)\bigg(\prod_{n=1}^k N_0\big(\nu_{\mathcal{D}_k^N(n)}\big) \mathcal{E}\big(\bx_{\mathcal{P}_k \mathcal{B}_k^N(n)}, \hOmega_{\mathcal{D}_k^N(n)}; \nu_{ \mathcal{D}_k^N(n)}\big) \delta\big(\omega_{\mathcal{P}_k\mathcal{B}_k^N(n)} - \nu_{\mathcal{D}_k^N(n)}\big) \bigg),
\end{align}
where $\mathcal{D}_k^N$ is another unordered $k-$element subset of  $\{1, 2 \dots N\}$ with $\bar{\mathcal{D}}^N_k$ being its complement, $\mathcal{P}_k$ is a $k-$ element permutation and $\mathcal{P}_k\mathcal{D}_k^N$ is a permutation of the $k-$element subset $\mathcal{D}_k^N$. Next, we substitute Eq.~\ref{eq:app_comm} into Eq.~\ref{eq:simp_smat} and use Eq.~\ref{eq:smat_heis_pic} to substitute for $a^\dagger(\hOmega, \nu; t_i)$ in terms of $a^\dagger(\hOmega, \nu; t_f)$. We note that due to the time-ordering operator, $\hat{a}^\dagger(\hOmega, \nu; t_f)$ annihilate $\bra{\text{vac}, g}$. With this simplification, we obtain Eq.~\ref{eq:smat_gen_exp} in the main text.

\section{Computing single and two-excitation Green's functions}
In this section, we outline the numerical procedures used for computing the single and two-excitation Green's function that are needed for calculating the single and two-excitation scattering matrices.
\subsection{Single-excitation Green's function}\label{app:single_ex_gfunc}
The single-excitation Green's function in time-domain, $\mathcal{G}(t; t')$, is given by
\begin{align}
 \mathcal{G}(t, s) = \bra{g, \text{vac}} \mathcal{T}\big[\sigma(t)\sigma^\dagger(s)\big]\ket{g, \text{vac}},
 \end{align}
where $\sigma(t)$ and $\sigma^\dagger(t)$ are the Heisenberg picture operators corresponding to the two-level system's lowering and raising operators. Since $\sigma(t)\ket{g, \text{vac}} = 0$ and $\bra{g, \text{vac}}\sigma^\dagger(t) = 0$ for all $t$, we obtain an alternative representation of the single-excitation Green's function
\begin{align}
\mathcal{G}(t, s) = \bra{e, \text{vac}} e^{-iH(t - s)} \ket{e, \text{vac}}\Theta(t \geq t'),
\end{align}
i.e.~it can be obtained by solving for the dynamics of an initially excited emitter emitting into the bosonic bath. To this end, we consider the system to initially be in the state $\ket{\psi(0)} = \sigma^\dagger \ket{G}$ --- using the noise operator representation of the bath, the following ansatz can be assumed for the state of the system
\begin{equation}
\ket{\psi(t)} =\bigg[ A_e(t)\sigma^\dagger + \int_{\bx \in \mathbb{R}^3} \int_{\omega \in \mathbb{R}^+} A_g(\bx, \omega; t) \phi^\dagger_\alpha(\bx, \omega)d^3\bx \ d\omega\bigg] \ket{G},
\end{equation} 
with $A_e(0) = 1$ and $A_g(\bx, \omega; 0) = 0$. From the Schroedinger's equation, it then immediately follows that
\begin{align}
i\frac{dA_e(t)}{dt} = \omega_0& A_e(t) +iV_0 \int_{\bx \in \mathbb{R}^3}\int_{\omega \in \mathbb{R}^+} \bigg(\frac{\omega^2 \alpha(\omega)}{\pi}\bigg)^{1/2} \text{G}_\alpha(\bx_d, \bx; \omega)A_g(\bx, \omega; t) d^3 \bx d\omega\label{eq:spic_eq_1}, \\
i\frac{dA_g(\bx, \omega; t)}{dt} &= \omega A_g(\bx, \omega; t) - iV_0 \bigg(\frac{\omega^2 \alpha(\omega)}{\pi}\bigg)^{1/2} \text{G}_\alpha^*(\bx_d, \bx; \omega) A_e(t).\label{eq:spic_eq_2}
\end{align}
Integrating the second equation, we obtain
\begin{equation}
A_g(\bx, \omega; t) = -V_0 \bigg(\frac{\omega^2 \alpha(\omega)}{\pi}\bigg)^{1/2} \text{G}_\alpha^*(\bx_d, \bx; \omega) \int_{0}^t A_e(t') e^{-i\omega(t -t')}dt'.
\end{equation}
Substituting this back into Eq.~\ref{eq:spic_eq_1} and using Eq.~\ref{eq:gfunc_prop_2}, we thus obtain
\begin{equation}\label{eq:single_ex_gfunc}
\frac{dA_e(t)}{dt} = -i\omega_0 A_e(t) + \frac{V_0^2}{\pi} \int_{t'=0}^t \int_{\omega=0}^{\infty} \text{Im}\big[\text{G}_\alpha(\bx_d, \bx_d; \omega)\big] e^{-i\omega (t - t')} A_e(t') dt' \ d\omega.
\end{equation}
We note that while the above analysis assumes a loss $\alpha(\omega)$ in the dielectric distribution under consideration, it can be taken to $0$ in Eq.~\ref{eq:single_ex_gfunc} since $\lim_{\alpha(\omega) \to 0}\text{Im}\big[\text{G}_\alpha(\bx_d, \bx_d; \omega)]$ exists.\\

\noindent\emph{Numerical solution}: For a given permittivity distribution $\varepsilon(\bx)$, Im$\big[G(\bx_d, \bx_d; \omega)\big] $ can be computed using various numerical methods (such as FDTD or FDFD). To compute $A_e(t)$, we approximate Im$\big[G(\bx_d, \bx_d; \omega)\big]$ as a sum of Lorentzians within a sufficiently large bandwidth around the emitter frequency
\begin{equation}
\text{Im}\big[\text{G}(\bx_d, \bx_d; \omega)\big] = -\sum_{n} \frac{p_n}{(\omega - \omega_n)^2 + \gamma_n^2}.
\end{equation}
Eq.~\ref{eq:single_ex_gfunc} can the be expressed as
\begin{align}\label{eq:lor_single_ex_1}
\frac{dA_e(t)}{dt} = -i\omega_0 A_e(t) - \frac{V_0^2}{\pi}\sum_{n} \xi_n(t),
\end{align}
where
\begin{align}
\xi_n(t) = \int_{t'=0}^{t}\int_0^\infty \frac{p_n e^{-i\omega(t - t')}}{(\omega - \omega_n)^2 + \gamma_n^2}A_n(t')\ dt' d\omega.
\end{align}
Assuming that $\omega_n / \gamma_n \gg 1$,
\begin{align}
\int_0^\infty \frac{p_n e^{-i\omega(t - t')}}{(\omega - \omega_n)^2 + \gamma_n^2} d\omega \approx \int_{-\infty}^\infty \frac{p_n e^{-i\omega(t - t')}}{(\omega - \omega_n)^2 + \gamma_n^2} d\omega = \frac{\pi p_n}{\gamma_n} e^{-i\omega_n(t - t')}e^{-\gamma_n|t- t'|},
\end{align}
which immediately yields     
\begin{align}
\xi_n(t) = \frac{\pi p_n}{\gamma_n} \int_0^{t} e^{-i\omega_n(t - t')}e^{-\gamma_n(t - t')} A_n(t') dt'.
\end{align}
A differential equation for $\xi_n(t)$ can be derived from this expression by differentiating it with respect to $t$
\begin{align}\label{eq:lor_single_ex_2}
\frac{d\xi_n(t)}{dt} = -(i\omega_n + \gamma_n) \xi_n(t)  + \frac{\pi p_n}{\gamma_n} A_e(t).
\end{align}
In frequency-domain scattering matrix calculations, the quantity of interest is the one-sided fourier transform of $A_e(t)$
\begin{align}
A_e(\omega) = \int_0^\infty A_e(t) e^{i\omega t}dt.
\end{align}
Taking the one-sided fourier transform of Eqs.~\ref{eq:lor_single_ex_1} and \ref{eq:lor_single_ex_2}, we obtain an expression for $A_e(\omega)$
\begin{align}
A_e(\omega) = \bigg[i(\omega_0 - \omega) + V_0^2 \sum_{n} \frac{p_n / \gamma_n}{i(\omega_n - \omega) + \gamma_n} \bigg]^{-1}.
\end{align}\\

\noindent\emph{Weisskopf-Wigner approximation}: For permittivity distribution $\varepsilon(\bx)$ with a frequency-response much broader than the line-width of the emitter, Eq.~\ref{eq:single_ex_gfunc} can be approximated using the Weisskopf-Wigner formalism
\begin{align}
\frac{dA_e(t)}{dt} = -\bigg(i\omega_0 + \frac{\gamma}{2}\bigg) A_e(t),
\end{align}
where $\gamma = -2V_0^2 \text{Im}\big[G(\bx_d, \bx_d; \omega_0)]$ is the spontaneous emission decay rate of the emitter into the inhomogeneous bath. This yields an exponentially decaying solution for $A_e(t)$ and a complex Lorentzian for $A_e(\omega)$
\begin{align}
A_e(t) = e^{-i\omega_0 t}e^{-\gamma t/2} \text{ and }A_e(\omega) = \frac{1}{i(\omega_0 - \omega) + \gamma / 2}.
\end{align}
\subsection{Two-excitation Green's function}\label{app:two_ex_gfunc}
The evaluation of the two-excitation Green's function for the two-level system can be done by following the procedure outlined in Ref.~\cite{shi2015multiphoton}. As an initial step, the two-level system is replaced with a boson with an onsite repulsion as described by the Hamiltonian
\begin{align}
H_\text{0} = \omega_0 \sigma^\dagger \sigma + \frac{U_0}{2}(\sigma^\dagger)^2 \sigma^2,
\end{align}
instead of $\omega_0 \sigma^\dagger \sigma$ and the commutation relation $[\sigma, \sigma^\dagger] = \sigma_z$ replaced by the bosonic commutation $[\sigma, \sigma^\dagger] = 1$. The two-level system is recovered from this model by taking $U_0 \to \infty$. Next, we decompose the full hamiltonian of the TLS coupled to the inhomogeneous bath, $H$, as
\begin{align}
H = H_\text{L} + H_\text{NL},
\end{align}
where $H_\text{L} = H_\text{bath} + H_\text{int} + \omega_0 \sigma^\dagger \sigma$ and $H_\text{NL} = U_0(\sigma^\dagger)^2 \sigma^2 / 2$. We point out that $H_\text{L}$ is a quadratic hamiltonian when expressed in terms of the bosonic annihilation operators of the bath and $\sigma$, and consequently its dynamics are completely captured by two-point correlations between the different bosonic operators. We now consider the time-ordered expectation $\bra{\text{vac}, g} \mathcal{T}\big[\sigma(t_1)\sigma(t_2) \sigma^\dagger(s_1) \sigma^\dagger(s_2)\big]\ket{\text{vac}, g}$ and rewrite it as
\begin{align}\label{eq:change_frame}
\bra{\text{vac}, g} \mathcal{T}\big[\sigma(t_1)\sigma(t_2)& \sigma^\dagger(s_1) \sigma^\dagger(s_2)\big]\ket{\text{vac}, g} =\nonumber \\ &\bra{\text{vac}, g} \mathcal{T}\bigg[\tilde{\sigma}(t_1)\tilde{\sigma}(t_2) \tilde{\sigma}^\dagger(s_1) \tilde{\sigma}^\dagger(s_2)\exp\bigg({-i\int_{-\infty}^\infty \tilde{H}_\text{NL}(t)dt}\bigg)\bigg]\ket{\text{vac}, g},
\end{align}
where, for an operator $O$, $\tilde{O}(t) = e^{iH_\text{L}t}Oe^{-iH_\text{L}t}$. The exponential in Eq.~\ref{eq:change_frame} can be expanded to obtain
\begin{align}\label{eq:dyson_series}
\bra{\text{vac}, g} \mathcal{T}\big[&\sigma(t_1)\sigma(t_2) \sigma^\dagger(s_1) \sigma^\dagger(s_2)\big]\ket{\text{vac}, g} = 
\bra{\text{vac}, g} \mathcal{T}\big[\tilde{\sigma}(t_1)\tilde{\sigma}(t_2) \tilde{\sigma}^\dagger(s_1)\tilde{ \sigma}^\dagger(s_2)\big]\ket{\text{vac}, g} +\nonumber\\ &\sum_{k=1}^\infty \int_{\tau_1, \tau_2 \dots \tau_k \in \mathbb{R}}\frac{1}{k!}\bigg(-\frac{iU_0}{2}\bigg)^k\bra{\text{vac}, g} \mathcal{T}\bigg[\tilde{\sigma}(t_1)\tilde{\sigma}(t_2) \tilde{\sigma}^\dagger(s_1)\tilde{ \sigma}^\dagger(s_2)\prod_{i=1}^k \big(\tilde{\sigma}^\dagger(\tau_k))^2\tilde{\sigma}^2(\tau_k)\bigg]\ket{\text{vac}, g} \prod_{i=1}^k d\tau_i.
\end{align}
Each of the time-ordered expectations in the summation in the above equation can be evaluated through an application of the Wick's theorem since the operators involved in the expectation are Heisenberg operators with respect to a quadratic Hamiltonian $H_\text{L}$
\begin{subequations}\label{eq:wick_theorem}
\begin{align}
&\bra{\text{vac}, g} \mathcal{T}\big[\tilde{\sigma}(t_1)\tilde{\sigma}(t_2) \tilde{\sigma}^\dagger(s_1)\tilde{ \sigma}^\dagger(s_2)\big]\ket{\text{vac}, g} = g(t_1; s_1) g(t_2; s_2) + g(t_1; s_2) g(t_2; s_1), \\
&\bra{\text{vac}, g} \mathcal{T}\bigg[\tilde{\sigma}(t_1)\tilde{\sigma}(t_2) \tilde{\sigma}^\dagger(s_1)\tilde{ \sigma}^\dagger(s_2)\prod_{i=1}^k \big(\tilde{\sigma}^\dagger(\tau_k))^2\tilde{\sigma}^2(\tau_k)\bigg]\ket{\text{vac}, g} = \nonumber \\
&\ \ \ \ \ \ \ \ \ \ \ \ 2^{k-1}\sum_{\mathcal{P}_k}g(t_1; \tau_{\mathcal{P}_k(1)}) g(t_2; \tau_{\mathcal{P}_k(1)}) g(s_1; \tau_{\mathcal{P}_k(k)})g(s_2; \tau_{\mathcal{P}_k(k)})\prod_{i=1}^{k-1} g^2(\tau_{\mathcal{P}_k(i)}; \tau_{\mathcal{P}_k(i + 1)}).
\end{align}
\end{subequations}
where $\mathcal{P}_k$ is a $k-$element permutation of $\{1, 2 \dots k\}$ and $g(t; s) = \bra{\text{vac}, g}\mathcal{T}[\tilde{\sigma}(t)\tilde{\sigma}^\dagger(s)]\ket{\text{vac}, g}$. We note that $H_\text{NL}$ does not impact the dynamics of the system within the single-excitation subspace, and consequently $g(t; s) = \bra{\text{vac}, g}\mathcal{T}[\sigma(t) \sigma^\dagger(s)]\ket{\text{vac}, g} = A_e(t - s) \Theta(t\geq s)$ where $A_e(t)$ is defined in appendix \ref{app:single_ex_gfunc}. Finally, we compute the two-excitation Green's function $\mathcal{G}(\omega_1, \omega_2; \nu_1, \nu_2)$ defined in Eq.~\ref{eq:two_ex_gfunc} by Fourier transforming the time-ordered expectation in Eq.~\ref{eq:change_frame} and using Eqs.~\ref{eq:dyson_series} and \ref{eq:wick_theorem}. We thus obtain
\begin{align}
\mathcal{G}^2(\vec{\omega}; \vec{\nu}) = \mathcal{G}^1(\omega_1; \nu_1)\mathcal{G}^2(\omega_2; \nu_2) +  \mathcal{G}^1(\omega_1; \nu_2)\mathcal{G}^2(\omega_2; \nu_1) + \mathcal{G}_C^2(\vec{\omega}; \vec{\nu}),
\end{align}
where $\mathcal{G}^1(\omega; \nu)$ is the single-excitation Green's function and $\mathcal{G}^2_C(\vec{\omega}; \vec{\nu})$ is the connected part of the two-excitation Green's function
\begin{align}
\mathcal{G}_C^2(\vec{\omega}; \vec{\nu}) = -i\pi \frac{U_0\mathcal{G}_0(\nu_1)\mathcal{G}_0(\nu_2)\mathcal{G}_0(\omega_1)\mathcal{G}_0(\omega_2)}{1 + iU_0 \Gamma(\omega_1 + \omega_2)} \delta(\omega_1 + \omega_2 - \nu_1 - \nu_2),
\end{align}
where $\mathcal{G}_0(\omega)$ is defined in Eq.~\ref{eq:single_ex_gfunc_value} and
\begin{align}\label{eq:conv_g0}
\Gamma(E) = \int_0^\infty A_e^2(t) e^{iE t} dt = \frac{1}{2\pi}\int_{-\infty}^\infty \mathcal{G}_0(E - \omega)\mathcal{G}_0(\omega) d\omega.
\end{align}
Finally, in order to obtain the result for a two-level system, we take the limit of $U_0 \to \infty$, which yields the following expression for $\mathcal{G}_C^2(\omega_1, \omega_2; \nu_1, \nu_2)$
\begin{align}\label{eq:connected_part_gfunc}
\mathcal{G}_C^2(\omega_1, \omega_2; \nu_1, \nu_2) = -\pi \frac{\mathcal{G}_0(\nu_1)\mathcal{G}_0(\nu_2) \mathcal{G}_0(\omega_1) \mathcal{G}_0(\omega_2)}{\Gamma(\omega_1 + \omega_2)} \delta(\omega_1 + \omega_2 - \nu_1 - \nu_2).
\end{align}
\emph{Weisskopf-Wigner approximation}: Within the Weisskopf-Wigner approximation,
\begin{align}
\mathcal{G}_0(\omega) = \frac{1}{i(\omega_0 - \omega) + \gamma / 2}.
\end{align}
Therefore, it follows from Eq.~\ref{eq:conv_g0} that
\begin{align}
\Gamma(E) = \frac{1}{i(2\omega_0 - E) + \gamma},
\end{align}
with which we obtain
\begin{align}
\mathcal{G}_C^2(\vec{\omega}; \vec{\nu}) = -\frac{i\pi(\omega_1 + \omega_2 - 2\omega_0 - i\gamma) \delta(\omega_1 + \omega_2 - \nu_1 - \nu_2)}{(\nu_1 - \omega_0 - i\gamma / 2)(\nu_2 - \omega_0 - i\gamma / 2)(\omega_1 - \omega_0 - i\gamma/2)(\omega_2 - \omega_0 - i\gamma/2)}.
\end{align}

\section{Extension to Markovian localized systems with general level structures}\label{app:mark_scat}
The results presented in Section \ref{sec:tls_scat} can easily be generalized to localized quantum systems with complex level structures (e.g.~$\Lambda$-systems, V-level systems etc.) under the Markovian approximation. We consider a localized system with Hamiltonian $H_s$ interacting with the inhomogeneous bosonic bath with an interaction Hamiltonian identical to that in Eq.~\ref{eq:int_hamil} with $\sigma$ being the operator through which the low-dimensional system couples to the bosonic bath. Since the Heisenberg equations of motion for the bath operators (Eq.~\ref{eq:heis_pic}) are independent of the level structure of the localized system, the $N-$particle scattering matrix element defined in Eq.~\ref{eq:def_n_ph_smat} can still be related to the Green's function of the low-dimensional system using Eq.~\ref{eq:smat_gen_exp_2}. In this section, we show that under the Markovian approximation, the low dimensional system's Green's functions are completely determined by the propagator corresponding to a dissipative (non-Hermitian) Hamiltonian defined entirely within the system's Hilbert space. This is an extension of a similar well known result in waveguide QED \cite{xu2015input, trivedi2018few, caneva2015quantum} to the case of an inhomogeneous high-dimensional bath.

For notational convenience and generality, we restrict our attention to evaluating the following time-ordered expectation value
\begin{align}
\mathcal{E}(t_1, t_2 \dots t_M) = \bra{\text{vac}, g} \mathcal{T}\bigg[\prod_{i=1}^M c_i(t_i)\bigg] \ket{\text{vac}, g},
\end{align}
where $c_i$ are operators defined on the localized system's Hilbert space and $\ket{g}$ is the ground state of the localized system. We note that the $N-$excitation Green's function defined in Eq.~\ref{eq:nex_gfunc} is simply the Fourier transform of such a time-ordered expectation with the system operators $c_i$ being either $\sigma$ or $\sigma^\dagger$. Choosing to describe the Hilbert space of the bath via the noise operators $\phi_\alpha(\bx, \omega)$, this expectation can be expressed as the following path integral \cite{weinberg1995quantum}
\begin{align}\label{eq:path_int_corr}
\mathcal{E}(t_1, t_2 \dots t_M) = \int_{\substack{\phi_\alpha(\bx, \omega; t_i) = 0 \\ \phi_\alpha(\bx, \omega; t_f) = 0}} \mathcal{D}[\phi_\alpha(\bx, \omega; t), \phi^*(\bx, \omega; t)]\mathcal{D}s\ \bigg[\prod_{i=1}^M c_i(t_i)\bigg]e^{-i(\mathcal{S}_\text{sys} + \mathcal{S}_\text{bath} + \mathcal{S}_\text{int})},
\end{align}
where $t_i \to -\infty$, $t_f \to \infty$, $\mathcal{D}[\phi_\alpha(\bx, \omega; t), \phi^*(\bx, \omega; t)]$ is the path integral measure over the Hilbert space of the bath, $\mathcal{D}s$ is the path integral measure over the Hilbert space of the localized system, $\mathcal{S}_\text{sys}$ is the classical action for the localized system from $t = t_i$ to $t = t_f$ and
\begin{subequations}
\begin{align}
&\mathcal{S}_\text{bath} = \int_{t = t_i}^{t_f}\int_{\bx\in\mathbb{R}^3} \int_{\omega \in \mathbb{R}^+}\bigg( \phi^*(\bx, \omega; t) \frac{\partial}{\partial t}\phi_\alpha(\bx, \omega; t) - \omega \big|\phi_\alpha(\bx, \omega; t)\big|^2\bigg)dt \ d^3 \bx\ d\omega,\\
&\mathcal{S}_\text{int} = -iV_0 \int_{t = t_i}^{t_f}\int_{\omega \in \mathbb{R}^+}\bigg( \psi_\alpha(\bx_d, \omega; t) \sigma^*(t) - \psi_\alpha^*(\bx_d, \omega; t)\sigma(t)\bigg)dt \ d\omega,
\end{align}
where
\begin{align}
\psi_\alpha(\bx, \omega; t) = \bigg(\frac{\omega^2 \alpha(\omega)}{\pi}\bigg)^{1/2}\int \text{G}_\alpha (\bx, \bx'; \omega) \phi_\alpha(\bx', \omega; t) d^3 \bx'.
\end{align}
\end{subequations}
We note that the integrand in Eq.~\ref{eq:path_int_corr} is gaussian in $\phi_\alpha(\bx, \omega; t)$ and consequently the integral with respect to $\phi_\alpha(\bx, \omega; t)$ can be evaluated exactly using the stationary phase method.  The stationary point of the integrand in Eq.~\ref{eq:path_int_corr} with respect to $\phi_\alpha(\bx, \omega; t)$ is given by the classical path $\phi_\text{cl}(\bx, \omega; t)$ that minimizes the action $\mathcal{S}_\text{bath} + \mathcal{S}_\text{int}$, which is governed by
\begin{align}
i\frac{\partial}{\partial t} {\phi}_\text{cl}(\bx, \omega; t) = \omega \phi_\text{cl}(\bx, \omega; t) + iV_0\bigg(\frac{\omega^2\alpha(\omega)}{\pi}\bigg)^{1/2} \text{G}_\alpha^*(\bx, \bx_d; \omega) \sigma(t).
\end{align}
Since we are interested in evaluating expectations in which the bath is in the vacuum state, $\phi_\text{cl}(\bx, \omega; t_i) = 0$ --- the above equation of motion can then be integrated to obtain
\begin{align}
\phi_\text{cl}(\bx, \omega; t) = V_0 \bigg(\frac{\omega^2 \alpha(\omega)}{\pi}\bigg)^{1/2}\text{G}_\alpha^*(\bx, \bx_d; \omega) \int_{t_i}^t \sigma(t') e^{-i\omega(t -t ')}dt'.
\end{align}
Next, we perform a change of variables in the path integral from $\phi_\alpha(\bx, \omega; t)$ to $\delta \phi_\alpha(\bx, \omega; t) = \phi_\text{cl}(\bx, \omega; t) + \delta \phi_\alpha(\bx, \omega; t)$. Since $\mathcal{S}_\text{bath} + \mathcal{S}_\text{int}$ is quadratic in $\phi_\alpha(\bx, \omega; t)$ and $\phi_\text{cl}(\bx, \omega; t)$ is a stationary point of this action, it is expressible as a quadratic in $\delta \phi_\alpha(\bx, \omega; t)$ without any linear terms in $\delta \phi_\alpha(\bx, \omega; t)$
\begin{align}
\mathcal{S}_\text{bath} + \mathcal{S}_\text{int}=\mathcal{S}_0 + \delta\mathcal{S},
\end{align} 
where
\begin{align}
\delta \mathcal{S} =  \int_{t=t_i}^{t_f} \int_{\bx \in \mathbb{R}^3}\int_{\omega \in \mathbb{R}^+} \bigg(\delta \phi^*(\bx, \omega; t) \frac{\partial}{\partial t}\delta \phi_\alpha(\bx, \omega; t) - \omega \big|\delta \phi_\alpha(\bx, \omega; t)\big|^2 \bigg)dt\ d\bx\ d\omega,
\end{align}
and $\mathcal{S}_0$ is the action $\mathcal{S}_\text{bath} + \mathcal{S}_\text{int}$ evaluated at $\phi_\text{cl}(\bx, \omega; t)$
\begin{align}
\mathcal{S}_0 &= -iV_0 \int_{t = t_i}^{t_f}\int_{\bx\in \mathbb{R}^3} \int_{\omega \in \mathbb{R}^+} \bigg(\frac{\omega^2 \alpha(\omega)}{\pi}\bigg)^{1/2}\text{G}_\alpha(\bx_d, \bx; \omega) \phi_\text{cl}(\bx, \omega; t)\sigma^*(t) dt \ d^3\bx\ d\omega, \nonumber \\
&= \frac{i V_0^2}{\pi} \int_{t=t_i}^{t_f} \int_{t' = t_i}^{t} \int_{\omega \in \mathbb{R}^+} \text{Im}\big[\text{G}(\bx_d, \bx_d; \omega)\big] \sigma^*(t)\sigma(t')e^{-i\omega (t - t')}  dt \ dt' \ d\omega,
\end{align}
wherein in the second step we have used the fact that $\lim_{\alpha(\omega) \to 0} \text{Im}[\text{G}_\alpha(\bx_d, \bx_d; \omega)]$ exists and have explicitly taken this limit. The expectation $\mathcal{E}(t_1, t_2 \dots t_N)$ then evaluates to
\begin{align}
\mathcal{E}(t_1, t_2 \dots t_M) = \mathcal{A}  \int \mathcal{D}s \ \bigg[ \prod_{i=1}^M c_i(t_i)\bigg] e^{-i\mathcal{S}_\text{eff}}
\end{align}
where
\begin{align}\label{eq:nonlocal_lag}
\mathcal{S}_\text{eff} = \mathcal{S}_\text{sys} + \frac{i V_0^2}{\pi} \int_{t=t_i}^{t_f} \int_{t' = t_i}^{t} \int_{\omega \in \mathbb{R}^+} \text{Im}\big[\text{G}(\bx_d, \bx_d; \omega)\big] \sigma^*(t)\sigma(t')e^{-i\omega (t - t')}  dt \ dt' \ d\omega,
\end{align}
and
\begin{align}
\mathcal{A} = \int_{\substack{\delta\phi_\alpha(\bx, \omega; t_i) = 0 \\ \delta \phi_\alpha(\bx, \omega; t_f) = 0}} \mathcal{D}[\delta\phi_\alpha(\bx, \omega; t), \delta\phi^*(\bx, \omega; t)] e^{-i\delta \mathcal{S}}. 
\end{align}
We immediately notice that $\mathcal{A}$ is just the expectation $\bra{\text{vac}} e^{-iH_\text{bath}(t_f - t_i)} \ket{\text{vac}}$ and thus evaluates to 1. This yields
\begin{align}\label{eq:exp_reduced}
\mathcal{E}(t_1, t_2 \dots t_M) = \int \mathcal{D}s \ \bigg[ \prod_{i=1}^M c_i(t_i)\bigg] e^{-i\mathcal{S}_\text{eff}}.
\end{align}
We note that the effective action $\mathcal{S}_\text{eff}$ is, in general, a nonlocal action consequently making the integral in Eq.~\ref{eq:exp_reduced} difficult to evaluate directly. However, within the Markovian approximation \cite{carmichael2009open}, this effective action becomes local. This immediately follows from Eq.~\ref{eq:nonlocal_lag} by noting that the Markovian approximation around a (system resonance) frequency $\omega_0$ approximates $\text{Im}[\text{G}(\bx_d, \bx_d; \omega)]$ by $\text{Im}[\text{G}(\bx_d, \bx_d; \omega_0)]$ and extends the integral with respect to $\omega$ from $-\infty$ to $\infty$ to obtain:
\begin{align}
\mathcal{S}_\text{eff} =\mathcal{S}_\text{sys} +  \frac{i\gamma}{2} \int_{t=t_i}^{t_f}  |\sigma(t)|^2 dt
\end{align}
where $\gamma = V_0^2 \text{Im}[\text{G}(\bx_d, \bx_d; \omega_0)]$. From this local lagrangian, we immediately obtain that the expectation in Eq.~\ref{eq:exp_reduced} can be evaluated by evolving the localized system with respect to a dissipative (non-Hermitian) Hamiltonian $H_\text{eff}$ defined by:
\begin{align}
H_\text{eff} = H_\text{sys} - \frac{i\gamma}{2}\sigma^\dagger \sigma
\end{align}
\end{document}